\documentclass[twocolumn]{aastex63}
\pdfoutput=1

\usepackage{hyperref}

\newcommand{\apg}  	{^{>}_{\sim}}
\newcommand{\apll}  	{^{<}_{\sim}}

\shorttitle{Multiband Photometry and Spectroscopy in the G165 Field}
\shortauthors{Pascale et al.}
\graphicspath{{./}{figures/}}
\received{Feb. 1, 2022}
\accepted{April 28, 2022}
\published{June 17 2022}
\submitjournal{ApJ}
\begin{document}

\title{Possible Ongoing Merger Discovered by Photometry and Spectroscopy in
 the Field of the Galaxy Cluster PLCK G165.7+67.0}

\author[0000-0002-2282-8795]{Massimo Pascale}
\affiliation{Department of Astronomy, University of California,
  501 Campbell Hall \#3411, Berkeley, CA 94720, USA}

\author[0000-0003-1625-8009]{Brenda L.~Frye}
\affiliation{Department of Astronomy/Steward Observatory, University of
  Arizona, 933 N. Cherry Avenue, Tucson, AZ 85721, USA}

\author[0000-0003-2091-8946]{Liang Dai}
\affiliation{Department of Physics, University of California, 366 Physics
  North MC 7300, Berkeley, CA. 94720, USA}

\author{Nicholas Foo}
\affiliation{Department of Astronomy/Steward Observatory, University of
  Arizona, 933 N. Cherry Avenue, Tucson, AZ 85721, USA}

\author[0000-0003-3658-6026]{Yujing Qin}
\affiliation{Department of Astronomy/Steward Observatory, University of
  Arizona, 933 N. Cherry Avenue, Tucson, AZ 85721, USA}

\author{Reagen Leimbach}
\affiliation{Department of Astronomy/Steward Observatory, University of
  Arizona, 933 N. Cherry Avenue, Tucson, AZ 85721, USA}

\author[0000-0002-7471-8934]{Adam Michael Bauer}
\affiliation{Department of Physics, University of Illinois at Urbana
  Champaign, 1110 West Green St., Loomis Laboratory, Urbana, IL 61801, USA}

\author[0000-0001-6870-8900]{Emiliano Merlin}
\affiliation{INAF - Osservatorio Astronomico di Roma, Via Frascati 33,
  I-00040 Monte Porzio Catone (RM), Italy}

\author[0000-0001-7410-7669]{Dan Coe}
\affiliation{STScI, 3700 San Martin Drive, Baltimore, MD 21218, USA}

\author[0000-0001-9065-3926]{Jose Diego}
\affiliation{FCA, Instituto de Fisica de Cantabria (UC-CSIC), Av.  de Los
  Castros s/n, E-39005 Santander, Spain}

\author[0000-0001-7592-7714]{Haojing Yan}
\affiliation{Department of Physics and Astronomy, University of Missouri,
  Columbia, MO 65211}

\author[0000-0002-0350-4488]{Adi Zitrin}
\affiliation{Physics Department, Ben-Gurion University of the Negev,
  P. O. Box 653, Be’er-Sheva, 8410501, Israel}

\author[0000-0003-3329-1337]{Seth H. Cohen}
\affiliation{School of Earth \& Space Exploration, Arizona State University,
  Tempe, AZ 85287-1404, USA}

\author[0000-0003-1949-7638]{Christopher J. Conselice}
\affiliation{Jodrell Bank Centre for Astrophysics, University of Manchester,
  Oxford Road, Manchester UK}

\author{Herv\'{e} Dole}
\affiliation{Universit\'{e} Paris-Saclay, Institut d’Astrophysique Spatiale,
  CNRS, bât 121, 91400 Orsay, France}

\author[0000-0001-5429-5762]{Kevin Harrington}
\affiliation{European Southern Observatory, Alonso de C´ordova 3107, Vitacura,
  Casilla 19001, Santiago de Chile, Chile}

\author[0000-0003-1268-5230]{Rolf A. Jansen}
\affiliation{School of Earth \& Space Exploration, Arizona State University,
  Tempe, AZ 85287-1404, USA}

\author{Patrick Kamieneski}
\affiliation{Department of Astronomy, University of Massachusetts at Amherst,
  Amherst, MA 01003, USA}

\author[0000-0001-8156-6281]{Rogier A. Windhorst}
\affiliation{School of Earth \& Space Exploration, Arizona State University,
  Tempe, AZ 85287-1404, USA}

\author[0000-0001-7095-7543]{Min S.~Yun}
\affiliation{Department of Astronomy, University of Massachusetts at Amherst,
  Amherst, MA 01003, USA}

\begin{abstract}

We present a detailed study of the {\it Planck}-selected binary galaxy
cluster PLCK~G165.7+67.0 (G165; $z$\,=\,0.348).  A multiband photometric
catalog is generated incorporating new imaging from the Large Binocular
Telescope/Large Binocular Camera and {\it Spitzer}/IRAC to existing imaging.
To cope with the different image characteristics, robust methods are applied
in the extraction of the matched-aperture photometry. Photometric redshifts
are estimated for 143 galaxies in the 4 arcmin$^{2}$ field of overlap
covered by these data. We confirm that strong lensing effects yield 30
images of 11 background galaxies, of which we contribute new photometric
redshift estimates for three image multiplicities.  These constraints enable
the construction of a revised lens model with a total mass of M$_{600 kpc}$\,=\,(2.36\,$\pm$\,0.23)\,$\times$\,10$^{14}$\,M$_{\odot}$. In parallel, new spectroscopy using MMT/Binospec and archival data contributes thirteen galaxies which meet our velocity and transverse radius criteria for
cluster membership.  The two cluster components have a pair-wise velocity of
$\apll$100~km~s$^{-1}$, favoring an orientation in the plane of the sky with
a transverse velocity of 100-1700~km~s$^{-1}$. At the same time, the
brightest cluster galaxy is offset in velocity from the systemic mean value, {suggesting dynamical disturbance}.
New LOFAR and VLA data uncover head-tail radio galaxies in the BCG and a large red galaxy in the
northeast {component. From the  orientation and alignment of the four radio trails, we infer that the two cluster components have already traversed each other, and are now exiting the cluster.}

\end{abstract}


\section{Introduction} \label{sec:intro}

Ever since the discovery that a giant arc in a cluster field is the strongly-lensed image of a background galaxy \citep{Soucail1987}, astronomers have come to
equate galaxy clusters with being powerful gravitational lenses. Strong
lensing distorts the images of the background galaxies into shapes which
trace out the gravitational potential of the cluster, in some cases
rendering the image of a single galaxy into multiple locations. The image
positions and redshifts of these image multiplicities place strong
constraints on the distribution of the lensing mass.  Starting at first by
incorporating the constraints from one image multiplicity to anchor the lens
model \citep{Franx1997, Frye1998}, to increasing that number up to dozens of
image multiplicities \citep{Kneib2004, Broadhurst2005}, the modeling schemes
grew over time into a precision science \citep[{\it e.\,g.,}][]{Jullo2007,
Zitrin2009, Zitrin2015, Kneib2011}.  Lens models can flag the high
magnification sightlines, offering windows into the distant universe at
$z$\,$\apg$\,10 \citep[{\it e.\,g.},][]{Coe2013}, as well as rare studies of
the interstellar medium within lensed sources at $z$\,$\apg$\,1 \citep[{\it
e.\,g.},][]{Frye2012,Sharma2018,Fujimoto2021,Nagy2021}.  Knowing the exact
placement of the critical curve enables searches at close angular
separations ($\apll$2\,$^{\prime \prime}$), resulting in detections of
compact sources with ultra-high magnification factors of several thousands
{\citep{Miralda1991, Venumadhav2017, Diego2018, Oguri2018, Windhorst2018,
Dai2018, Dai2020, Dai2020b, Vanzella2020}}, and yielding the discovery of
individual stars at cosmological distances known as caustic transients
\citep{Kelly2018,Kaurov2019,Chen2019,Welch2022}.

Obtaining the full entourage of image constraints is a challenge owing to
limitations of high-quality and high-resolution observations. It is natural to ask how image constraints affect the
construction of a reliable lens model, by which it is meant that the source
plane positions, redshifts, and/or magnification factors are accurately
recovered. 

In one study, \cite{Ghosh2020} analyzed simulations of clusters
and found that introducing additional image constraints generally increased
the accuracy of the mass map. To quantify the systematics of image
constraints, \citet{Johnson2016} analyzed the strong lensing model of a
single simulation of a single cluster hundreds of times, differing only the
amounts and types of lensing evidence. They found that a reliable lens model
is obtained for lensing constraints consisting of $\geq$\,10 image
multiplicities, of which $\geq$\,5 had measured spectroscopic
redshifts. Failing to input any spectroscopic redshift information was found
to provide a poor fit relative to the fiducial lens model.  They further
noted that supplying photometric redshifts improved lens model reliability,
a result that is especially impactful when photometric estimation is the
only practicable option. These considerations have motivated this study of
multi-band imaging and of spectroscopy in the field of the massive lensing
cluster PLCK~G165.7+67.0 (G165).

G165 was discovered using data from the {\it Planck} survey
\citep{PlanckCollaboration2018} and {\it Herschel Space Observatory}
\citep{Pilbratt2010} in a census of cluster-scale structures by its
rest-frame far-infrared colors and {\it not} by the Sunyaev-Z'eldovich (SZ)
effect \citep{PlanckCollaboration2015,Canameras2015}. Interestingly, G165
yields only an upper limit on the mass from the {\it Planck} Compton-Y map,
and is a low-luminosity X-ray source {despite its bimodal configuration}
\cite[][hereafter F19]{Frye2019}. By contrast, Hubble Space Telescope ({\it
HST}) imaging uncovers rich lensing evidence.  Eleven image multiplicities
consisting of 30 images are identified,

{ many of which are `caustic-crossing'
arcs that constrain the position of the critical curve. The positions
and redshift information of these images are used to construct the lens model, and to estimate its mass ($10^{14}M_{\odot}$). The high measured mass and low X-ray luminosity mean that G165 may be participating in a major cluster-cluster merger in its initial infall, or in a relatively common but more distant cluster-cluster interaction. The current lens model, however,
lacks the precision necessary to carry out this further work.} This is not so
surprising given that there are no published photometric redshifts, and a
spectroscopic redshift has been measured for only one member of one image
system, Arc 1a (initially in \citet{Canameras2015}, with
uncertainties in \citet{Harrington2016}, and later with multiple CO/[CI]
emission lines in \citet{Canameras2018} and \citet{Harrington2021}). {Moreover, the cluster
member catalog is estimated from a red-sequence fit guided by only a few
spectroscopically confirmed cluster members. Redshift information on both
the cluster galaxies and the image systems will significantly improve the
precision of the lens model.}

{Here we present new imaging and spectroscopy in the G165 field}. These data
are combined with archival imaging so that the optical through near-infrared
spectral energy distributions (SEDs) and photometric redshifts can be fit.
Robust techniques are applied to assemble the multi-band photometric catalog
from these somewhat disparate imaging data sets by applying the
template-fitting approach of \texttt{T-PHOT} \citep{Merlin2015,Merlin2016a}. The
analysis yields new photometric redshifts for multiple members of three image
systems and numerous cluster galaxies. {These redshifts provide constraints on the lens model and hence the lensing mass distribution.} We also acquired new
spectroscopy and new radio maps which enables us to better understand the
cluster kinematics and dynamics.

The paper is organized as follows: in \S\ref{sec:obs} we present the new
imaging and spectroscopy.  The algorithms used to prepare the detection
image for photometry are given in \S\ref{sec:prep}, and the detailed methods
regarding object detection and incorporation of the new and archival imaging
data sets are described in \S\ref{sec:phot}. The photometric imaging results such as the SED fits and the estimation of photometric redshifts are presented in \S\ref{sec:phot_results}. The spectroscopic results, such as the identification of new cluster members are
given \S\ref{sec:spec_results}. The revised lens model based on the
newly-obtained lensing constraints is presented in \S\ref{sec:LTM}, with an
analysis of the high-redshift population and Arc~1. Clues as to the cluster's evolutionary state drawn from the multi-wavelength data including new VLA and LOFAR imaging are discussed in \S\ref{sec:cluster_ev}.  The summary and conclusions appear in \S\ref{sec:end}.  We use the AB magnitude system
throughout this paper and we assume a $\Lambda$CDM cosmology with
$H_0$\,=\,67~km~s$^{-1}$\,Mpc$^{-1}$, $\Omega_{m,0}$\,=\,0.32, and
$\Omega_{\Lambda,0}$\,=\,0.68 \citep{PlanckCollaboration2018}.

\section{Observations and Reductions} \label{sec:obs}

\subsection{{Optical/Infrared Imaging}}
\subsubsection{Large Binocular Telescope}

Observations were obtained with the Large Binocular Telescope (LBT) Large
Binocular Camera (LBC) in 2018 January 20 (2018A, PI: Frye).  The imaging
includes $g$- (68 minutes total exposure), and in $i$-bands (43 minutes) in
2018 January 20 (2018A, PI: Frye) at a native plate scale of
0.2255$^{\prime\prime}$\,pixel$^{-1}$.  We refer to Table \ref{tab:1} for
the observing details.  The image reductions were carried out using Theli
(version 2.10.5), which is a general astronomical image reduction pipeline
operated through a web-based interface \citep{Erben2005, Schirmer2013}.
Details regarding the Theli application in this work have been documented
separately and made publicly
available\footnote{www.cloudynights.com/topic/679713-write-up-for-inexperienced-theli-users/
\href{www.cloudynights.com/topic/679713-write-up-for-inexperienced-theli-users/}{}}.

\begin{figure}[ht]
	\centering\includegraphics[scale =0.36]{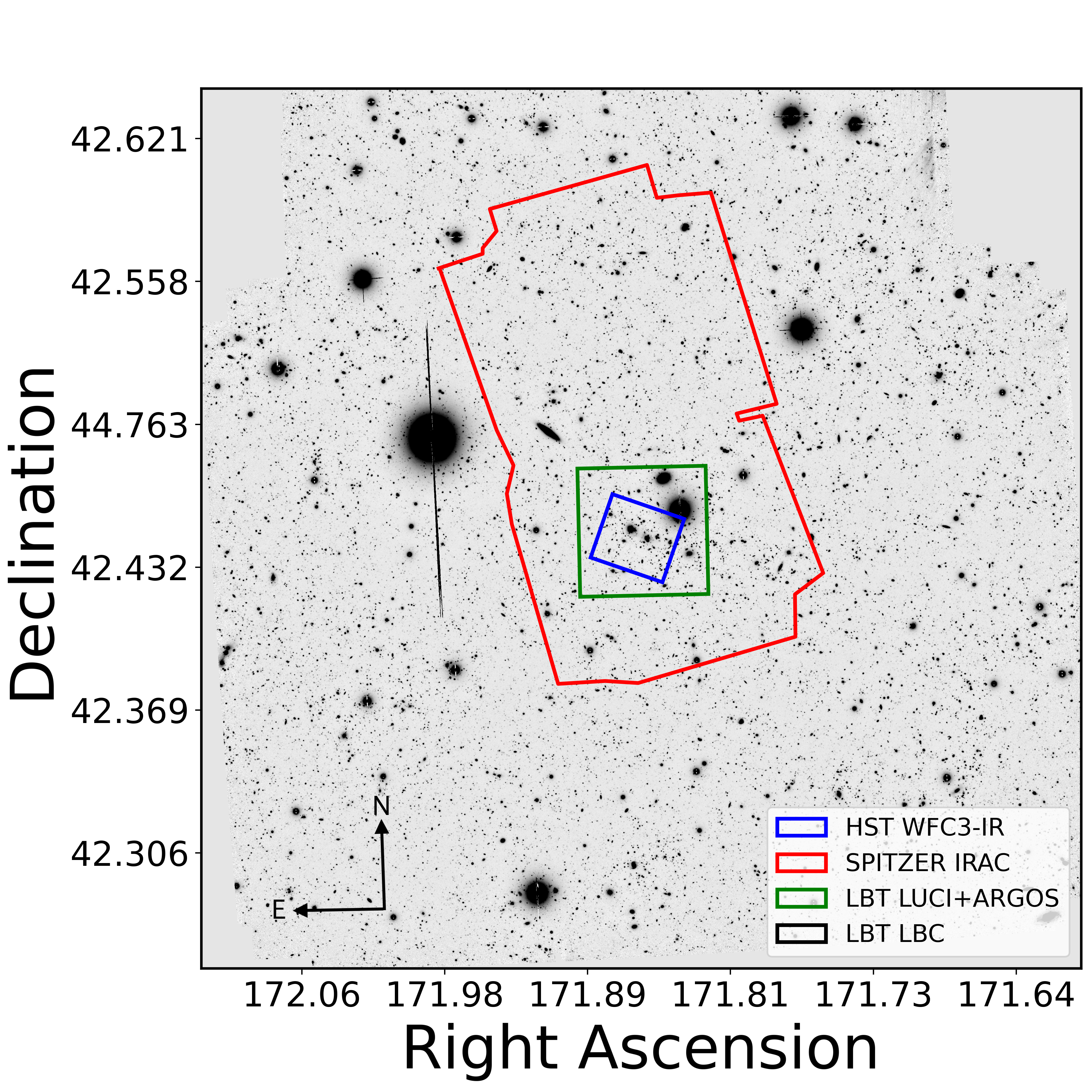}
	\caption{Footprint of the four instrument mosaics which make up the
7 filters contained in this study. The LBT/LBC field-of-view is depicted by
the x- and y-axes, equating to 25$^{\prime}$\,$\times$\,27$^{\prime}$ (black
frame). The fields-of-view covered by {\it Spitzer}/IRAC (red), LBT
LUCI+ARGOS (green), and {\it HST} WFC3-IR (blue) are also depicted.  The
{\it HST} WFC3-IR filters limit the FOV for which photometric redshifts can
be estimated to the central $\sim$2 arcminutes of the cluster, which equates
to $\sim$600 kpc at the mean cluster redshift of 0.35.}
	\label{fig:foot}
\end{figure}

Briefly, we performed initial calibrations of the bias-subtraction and
flatfield corrections in the usual way.  To correct for fringing, a
superflat was generated by masking out the bad pixels and bright sources
over subsets of contiguous science frames, median-combining them, and then
dividing the superflat into the data. To compute the astrometric solution,
the SCAMP software \citep{Bertin2002, Bertin2006} was run through the Theli
wrapper, selecting the Initial GAIA Source Catalog (IGSC) as the astrometric
reference \citep{GAIA2016}.  Visuals and intermediate data products were
produced as sanity checks to ensure that the dither pattern was recognized,
the distortion pattern was measured, and the positional residuals were low
and evenly distributed about zero.  Following the initial frame calibration
and registration, the background was modeled using Theli's internal
pipeline.  Finally, the individual frames were co-added using SWarp
\citep{Bertin2010} run from within the Theli GUI. Figure \ref{fig:foot}
depicts the $i$-band image of the full
25$^{\prime}$\,$\times$\,27$^{\prime}$ field-of-view (FOV) of LBT/LBC
(outermost black frame), which reaches 3-$\sigma$ limiting magnitudes of
$g_{AB}$ = 25.42 mag, and $i_{AB}$ = 24.67 mag.

\begin{figure}[h]
	\centering\includegraphics[scale =0.63]{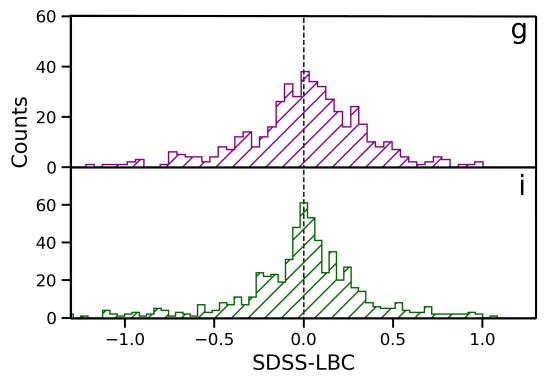}
	\caption{Histogram of the SDSS photometry relative to our LBT/LBC
photometry.  We measure mean systematic offsets within the central
5$^{\prime}$\,x\,5$^{\prime}$ FOV of 0.37 and 0.29 mag in $g$- and
$i$-bands, respectively. The magnitude difference histograms reflect the
distributions after the correction has already been made. The corrected
photometric zeropoints are recorded in Table \ref{tab:1}. We refer also to
\S4.6.1 for details regarding an independent test of the zeropoints.}
	\label{fig:sdss}
\end{figure}

To test the absolute $gi$ LBT/LBC photometry, the positions of the sources
in the central 5\,$\times$\,5 arcmin FOV were matched with their
counterparts in the Sloan Digital Sky Survey (SDSS) DR16 catalog
\citep{Ahumada2020}. On computing the magnitude differences of
$g_{SDSS}-g_{LBC}$ and $i_{SDSS}-i_{LBC}$, and making a 3-$\sigma$ clip to
remove outliers, we measure mean offsets of 0.37 and 0.29 mag for $g$ and
$i$, respectively.  This means that our LBT/LBC photometry was
systematically brighter by those amounts.  Figure~\ref{fig:sdss} shows the
magnitude difference histograms following the zeropoint offset correction
that minimizes the mean difference to zero.  The adopted photometric
zeropoints are recorded in Table~\ref{tab:1}, and are used to extract the
photometry.  In sum, these images are deep, making them valuable for this
study, and giving them leverage as veto bands for the planned high-redshift
galaxy searches using {Prime Extragalactic Areas for Reionization and Lensing Science (PEARLS; program
\#1176).}

\begin{deluxetable}{ccccc}
\tablecaption{The Sample:  Observing Details}
\tablecolumns{4}
\tablehead{
 \colhead{Filter$^a$} & \colhead{Exp.\,(s)} & \colhead{$m_{lim}^b$} &
 \colhead{FWHM ($^{\prime \prime}$)$^c$} & \colhead{$m_{zp}^d$}\\
}
\startdata
$g$      & 4090 & 25.42 & 1.37 & 27.57\,+\,0.25\\
$i$      & 2600 & 24.67 & 1.07 & 28.31\,+\,0.32\\
$F110W$  & 2664 & 28.94 & 0.13 & 26.82\,+\,0.05\\
$F160W$  & 2556 & 27.97 & 0.15 & 25.95\,+\,0.00\\
$K$      & 120  & 24.07 & 0.29 & 25.74\,+\,0.18\\
Ch1[3.6] & 1200 & 23.31 & 1.95 & 21.59\,+\,0.21\\
Ch2[4.5] & 1200 & 23.33 & 2.02 & 21.58\,+\,0.14\\
\enddata
\tablenotetext{a}{The filter images were acquired as follows: $gi$-bands
using LBT/LBC, $F110W$ and $F160W$ bands using {\it HST} WFC3-IR, $K$-band
using LBT/LUCI+ARGOS, and Channel 1 at 3.6\,$\mu$m and Channel 2 at
4.5\,$\mu$m using {\it Spitzer}/IRAC.}
\tablenotetext{b}{The $3\sigma$ limiting magnitude is estimated by placing
apertures of twice the FWHM of the PSF in empty regions of sky.}
\tablenotetext{c}{The FWHM of the PSF is reported for each filter. We
measure PSF FWHM by fitting a Gaussian to the PSFs measured in \S4.3.}
\tablenotetext{d}{ Zeropoint magnitudes are given as the quoted instrumental
magnitude plus the observed zeropoint corrected in \S4.6.1.}
  \label{tab:1}
\end{deluxetable}

The LBT Advanced Rayleigh Ground layer adaptive Optics System (ARGOS;
\cite{Rabien2019}) operates through the existing LUCI
instrumentation. LUCI+ARGOS corrects the atmosphere for ground-layer
distortions via multiple artificial stars which are projected by laser beams
mounted on each of the two 8.4~m apertures. Observations of the G165 field
were obtained in the $K$-band in 2016 December 16 (2016B; PI: Frye).  We
acquired the LUCI+ARGOS imaging in monocular mode with native plate scale of
0.12$^{\prime \prime}$\,pixel$^{-1}$. The observing details are recorded in
Table~\ref{tab:1}, and the image reduction are presented in F19 and in
\citet{Rabien2019}. In the reduced images, we measure a value for the point
source FWHM\,$\approx$\,0.29$^{\prime \prime}$.  This angular resolution is
comparable to {\it HST} within a factor of two, effectively extending the
red wavelength reach of {\it HST}.  We performed our own photometry on these
data, using the instrumental zeropoint AB magnitude drawn from Table 9 of
the LBT LUCI users
manual\footnote{\url{https://lsw.uni-heidelberg.de/users/jheidt/LBT\_links/LUCI\_UserMan.pdf}}.

\subsubsection{Hubble Space Telescope}
Observations were obtained between 2015 December and 2016 July as a part of
an {\it HST} WFC3-IR program (Cy23, GO-14223, PI: Frye). Exposures were
taken in the $F110W$ and $F160W$ passbands (0.13$^{\prime
\prime}$\,pixel$^{-1}$), reaching 3-$\sigma$ point source limiting
magnitudes of 28.94 and 27.97 AB mag, respectively.  Details of the
observations are summarized in Table~\ref{tab:1}, and a description of the
calibration and image reductions can be found in F19. WFC3-IR $F160W$ is
used as the detection image or High Resolution Image (HRI) for the matched
aperture photometry performed in this study.

\subsubsection{Spitzer}
{\it Spitzer} Infrared Camera (IRAC) 3.6 and 4.5~$\mu$m imaging was taken in
2016 as a part of a larger program (Cy13, GO-13024, PI: Yan).  The observing
details can be found in Table~\ref{tab:1}, and the image reduction is
described elsewhere \citep{Griffiths2018}. Images of the central region, and
of the multiple images of the Dusty Star Forming Galaxy (DSFG) Arc~1 that is singularly detected using the {\it
Planck} telescope, appear in F19. The {\it Spitzer} data have a native plate
scale of 1.22$^{\prime \prime}$\,pixel$^{-1}$ that is larger than the other
filters used in this study.  Yet these data are important for their role in
settling photometric redshift ambiguities, particularly for distinguishing
between natural continuum breaks such as the 4000~\AA \ and Balmer breaks
and lower redshift sources that are intrinsically dusty. We extract the
photometry for the O/IR imaging by the \texttt{T-PHOT} method, as is
described in \S4.5.

\subsection{VLA and LOFAR}
Karl G.~Jansky Very Large Array (VLA) 6 GHz imaging was acquired as part of
a larger program (18A-399, PI: P. Kamieneski). Details of the image
reductions will appear in an upcoming paper (Kamieneski et al.~2022, in
preparation). Briefly, the G165 field was observed with C-band ($4-8$ GHz)
with full polarization in a 3-hour track to improve $uv$-coverage, amounting
to a total of 1.5 hours of on-source integration time. Baselines for this
A-configuration observation ranged from 0.68 km to 36.4 km, resulting in a
natural-weighted synthesized beam size of $0.65\arcsec \times 0.37\arcsec$
at a position angle of 72$^{\circ}$ and a maximum recoverable scale of
$8.9\arcsec$. The sensitivity is measured to be 2.7 $\mu$Jy~beam$^{-1}$
across the approximately 4 GHz of effective bandwidth. Data calibration and
reduction were performed with the Common Astronomy Software Applications
(CASA) package \citep{McMullin2007}, and the VLA Calibration pipeline. The
radio maps uncover two head-tail radio galaxies which are discussed in
\S8.2.

Low Frequency Array (LOFAR) radio observations were acquired on 2016 March
22 through a single object request (PIs: Lehnert, Dole \& Frye), as part of
the Two-meter Sky Survey \citep[LoTSS;][]{Shimwell2017,Shimwell2019}. The
FITS image was delivered on 2019 February 19, following on-site calibration
by the Default Preprocessing Pipeline \citep[DPPP;][]{vanDiepen2018}.  These
low-frequency (120$-$168~MHz) data have a typical angaular resolution of
$\sim$6$^{\prime \prime}$ and a sensitivity of $\sim$100
$\mu$Jy~beam$^{-1}$.  Large-scale radio emission is detected over a
$\sim$500~kpc scale that potentially arises from radio jets and/or electrons
which are re-energized as a consequence of large-scale turbulence.

\begin{figure*}[t!]
\centering\includegraphics[scale=0.165]{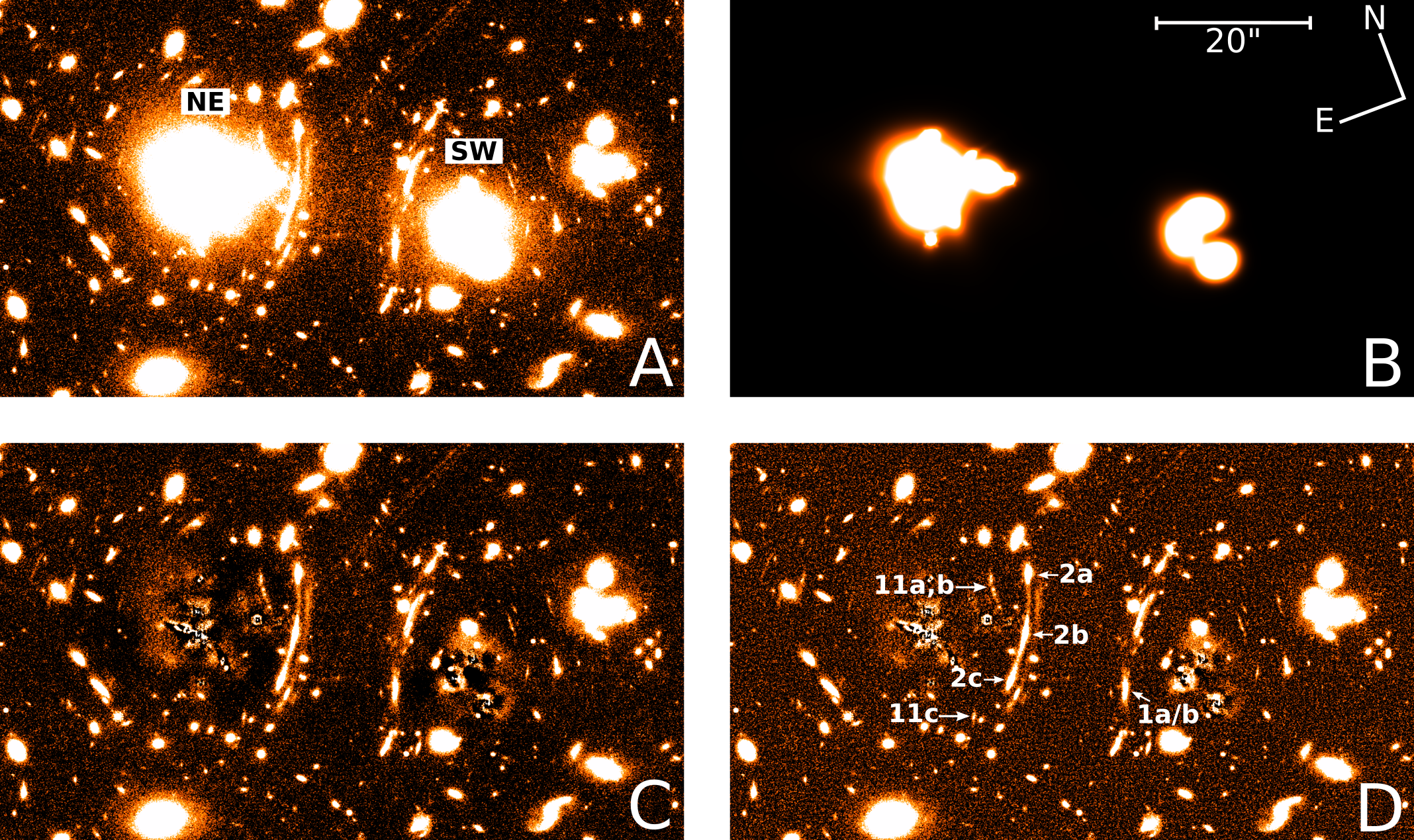}
\caption{Preparation of the HRI $F160W$, as described in \S3.  The panels
represent each stage in order of operation: (Panel A) the original reduced
$F160W$ image; (Panel B) the best fit \texttt{Galfit} models for the 13
central galaxies and ICL discussed in \S3.1 and \S3.2; (Panel C) the model
subtracted $F160W$ image; (Panel D) the median corrected image, which
significantly flattens the residual over-subtractions discussed in
\S3.3. The two sides of the cluster are referred to as the northeast (NE)
and southwest (SW) components, as labelled. The image sets for which we
report new photometric redshift estimates are also labeled and discussed
separately in \S4.6.2.}
\label{fig:imprep}
\end{figure*}

\subsection{Spectroscopy}

New multi-slit spectroscopy was obtained on 2019 February 8 using
MMT/Binospec (PI: Frye, 2019A).  To maximize the redshift search space, we
selected the 270 lines mm$^{-1}$ grating, which covers a wavelength range of
5500\,\AA\ about the central wavelength of each slitlet at a dispersion of
1.3 \AA~pix$^{-1}$.  We observed the central region of the G165 cluster in a
single pointing consisting of two 8$^{\prime}$\,$\times$\,15$^{\prime}$
fields at a separation of 3.2 arcminutes.  A total of 88 bright
($i_{AB}$\,$<$\,22) galaxies populated the slitmasks.  The observations were
composed of 6\,$\times$\,1200~s exposures, and were acquired under
relatively-stable seeing conditions of $\sim$1$^{\prime \prime}$ as measured
off of isolated sky lines.  This result sufficed for our science goal to
measure spectroscopic redshifts given the 1$^{\prime \prime}$ slitwidth and
the relatively-bright magnitudes of our targets. The data were reduced using
the observatory pipeline, which performed the bias correction,
flat-fielding, wavelength calibration, relative flux correction,
co-addition, and extraction to 1D
spectra\footnote{\url{https://bitbucket.org/chil\_sai/binospec}}. Although
redshift-fitting software was available, we opted to measure the
spectroscopic redshifts of the coadded 1D spectroscopy in
IDL\footnote{Interactive Data Language;
\url{https://www.l3harrisgeospatial.com/Software-Technology/IDL}} by way of
our own software. The SPEC task has a library of spectroscopic features and
a reference sky spectrum extracted from the pre-sky-subtracted data, and is
described elsewhere \citep[][and references therein]{Frye2012}. Our catalog
contains the secure spectroscopic redshifts for 86 of 88 sources. Of these,
we find eight new galaxies which meet our criteria for cluster membership as
defined in \S6.2.

\section{Image Preparation for Photometry of the HRI} \label{sec:prep}

\subsection{Motivation}
Limited to a 7-band filter suite, the photometry in each band is impactful
for making satisfactory SED fits. We perform the multiband photometry using
the prior-based software package \texttt{T-PHOT}, which is designed to do
photometry across images drawn from different observing facilities and/or
instrument modes, and to different field depths. A single HRI is designated
with which to extract the master object catalog.  To retain its information,
priors on object morphology are derived from the HRI which are then enforced
onto the images in the six other filters or Low Resolution Images (LRIs).

\texttt{T-PHOT} enlists two types of priors to generate a model of the
LRI. The `real' prior consists of cutouts from the HRI which are informed by
the photometry, and is especially sensitive to background sources of
light. To approximate this background, we carefully model both the ICL and
the bright galaxy light components, where the latter component becomes the
`analytic' prior. Our approach follows closely the methods of the ASTRODEEP
catalogs of the Frontier clusters \citep{Merlin2016a,Castellano2016}, and we
refer also to the companion flowchart in \cite[][their
Figure~1]{Merlin2016a}.  Once obtained, the background model is refined by
an iterative process, and then subtracted off of the HRI. The
background-subtracted HRI is further corrected for any residual light, as is
described below.

\subsection{{\it Galfit} Setup}
The background modeling algorithms used in this study make extensive use of
the \texttt{Galfit} tool \citep[version 3;][]{Peng2010}.  \texttt{Galfit}
operates on an input science image in units of counts, and a sigma image,
which is a map of the standard deviation of the science image.  The sigma
image can be generated externally, by using the weight files outputted from
the {\it HST} MultiDrizzle task following \cite{Casertano2000}, and also
internally by inputting the values from the image headers into
\texttt{Galfit}. To test the relative merits, we ran \texttt{Galfit} on a
``sky" object consisting of a rectangular isolated region of sky.  We found
the reduced $\chi^2$ to favor the internal approach, as the weight images
underestimated the actual noise in the complex cluster scene, and adopt this
method for generating the input sigma image for all of our \texttt{Galfit}
modeling.

\subsection{Initial Background Model Fits}
The background consists of two components: (1) the intracluster light (ICL),
and (2) the light from individual galaxies.  We constructed an initial model
for the ICL component by masking out all pixels in the HRI greater than
$8\sigma_{sky}$, where $\sigma_{sky}$ is estimated by sampling the sky in
three small and statistically significant patches of the image separated
from known sources. We then select the modified Ferrer profile included in
\texttt{Galfit}, which is prescribed in \cite{Merlin2016a} and
\cite{Giallongo2014} due to its relatively flat core and freedom to change
the sharpness of its truncation when compared to similar S\'{e}rsic
profiles. As expected, a combination of two modified Ferrer profiles
provided the best fit to the two light peaks visible in the data (Panel A of
Figure~\ref{fig:imprep}). We note that while parameters like bending modes,
diskiness/boxiness, and other profiles ({\it e.\,g.}, S\'{e}rsic) were
tried, they did not improve the results and hence were discarded. The
best-fit ICL model was then subtracted from the original image.

Our model for the cluster galaxy light contains the brightest cluster member
in the field, the Brightest Cluster Galaxy (BCG), and five other dominant
central cluster members or Large Red Galaxies {(LRGs)} and their nearest
satellites, to constitute a set of 13 galaxies. The inclusion of the
satellite galaxies enables a better fit to the galaxy light by our
\texttt{Galfit} approach.  These bright galaxies typically require more than
one \texttt{Galfit} profile to best model their overall brightness
profile. Simultaneously fitting many \texttt{Galfit} profiles, however, can
give rise to degenerate solutions, ultimately preventing \texttt{Galfit}
from converging to a best-fit solution. We instead adopt an iterative
approach, beginning with only a single S\'{e}rsic profile.  For this initial
fit, we make use of the software package \texttt{Galapagos}, which contains
the desired functionality by performing separate \texttt{Galfit} fits for
all objects at the positions given in SExtractor \citep{Barden2012}.  We
found that using a single profile per galaxy in this initial fitting stage
helps to prevent degeneracies in the subsequent two-profile refinement
fitting stage, as is described below.

We iterate on the initial galaxy light model by introducing an additional
S\'{e}rsic profile, such that each galaxy model is made up of two S\'{e}rsic
profiles. Following the prescription of \citet{Merlin2016a}, we use the best
fit parameters of the initial models to place constraints on the parameters
of the refinement model. Some individual galaxies required even further
refinement, with errors in the position angle and/or axis ratios evident by
a visual comparison between the model and image. We tighten up the fitting
constraints on an object by object basis, iterating as needed until a
solution is reached that, upon visual inspection, produces reasonably flat
residuals.

We then return to the ICL model and iterate on it by applying the results of
the revised galaxy light model as a fixed component in \texttt{Galfit},
while both ICL Ferrer components from the ICL initial fit are left
free. This step allows the ICL fit to be readjusted relative to the new
bright galaxy model fits, thereby superseding the simple 8$\sigma$ mask that
was initially enforced in \S3.1. The only other constraint placed on the ICL
components is a position center good to $\pm1$ pix from the initial value in
each coordinate.  The results did not vary much from the initial ICL fit,
providing a check on the quality of our overall fit. The resulting image
depicting the combined background light appears in Panel B of Figure
\ref{fig:imprep}.
\vspace{10mm}

\subsection{Residual Corrections}
Subtracting the background image from the data leaves regions of negative
flux. To alleviate the impact of these image residuals, we follow the
prescription in \cite{Merlin2016a}.  The procedure is to run SExtractor on
the sky surrounding each bright galaxy to estimate the sky RMS,
$\sigma_{SE}$. Next, the image is median-filtered using the PyRAF task
\texttt{median}, imposing a 1$^{\prime \prime}$\,x\,1$^{\prime \prime}$
window and excluding all pixels $>$\,1$\sigma_{SE}$ and neighboring pixels.
Finally, the median-filtered image is subtracted from the residual of the
original image. The resulting image has the expected flatter background
adequate for doing photometry (panel D of Figure \ref{fig:imprep}). Since
the photon noise is retained in the residual image, and the \texttt{Galfit}
models themselves contain no noise, we do not find it necessary to adjust
the $F160W$ RMS map in the subtracted image.

\section{Photometry} \label{sec:phot}

\subsection{Overview}
We describe below our measurement of the photometry in the {\it HST} filters
using SExtractor, and in the other LRIs using \texttt{T-PHOT}.  To achieve a
more realistic model for each LRI, \texttt{T-PHOT} is informed by the two
different priors described in \S3.1.  In what follows, the real prior is
obtained from the SExtractor catalog (\S4.2), and the analytical prior is
taken from the bright galaxy models (\S3.2).  The two priors are stitched
together into a single 2D image which is referred to as a collage.  This
collage is subsequently degraded to match the PSF, and then scaled to match
the flux of each LRI.  This final step in the process yields the photometry
for each LRI.

\subsection{Object Detection}
The {\it HST} WFC3-IR $F160W$ image is our object detection reference, or
HRI.  We perform the photometry using \texttt{SExtractor} \citep{Bertin1996}
by implementing a two-step \texttt{HOT+COLD} method according to the
prescription in \cite{Galametz2013}. In this prescription, the \texttt{COLD}
parameter set is tuned to the detection of bright, extended galaxies, while
the \texttt{HOT} parameter set is optimized to detect the fainter galaxies
not included in the preceding \texttt{COLD} mode run. We choose to match the
parameters of \cite{Galametz2013} rather than \cite{Merlin2016a} given the
limited depth of our HRI, as the \texttt{HOT} mode run in \cite{Merlin2016a}
is significantly more aggressive than \cite{Galametz2013}, producing many
more spurious sources without additional true sources in our case.
\begin{figure}[h]
	\centering\includegraphics[scale =0.55]{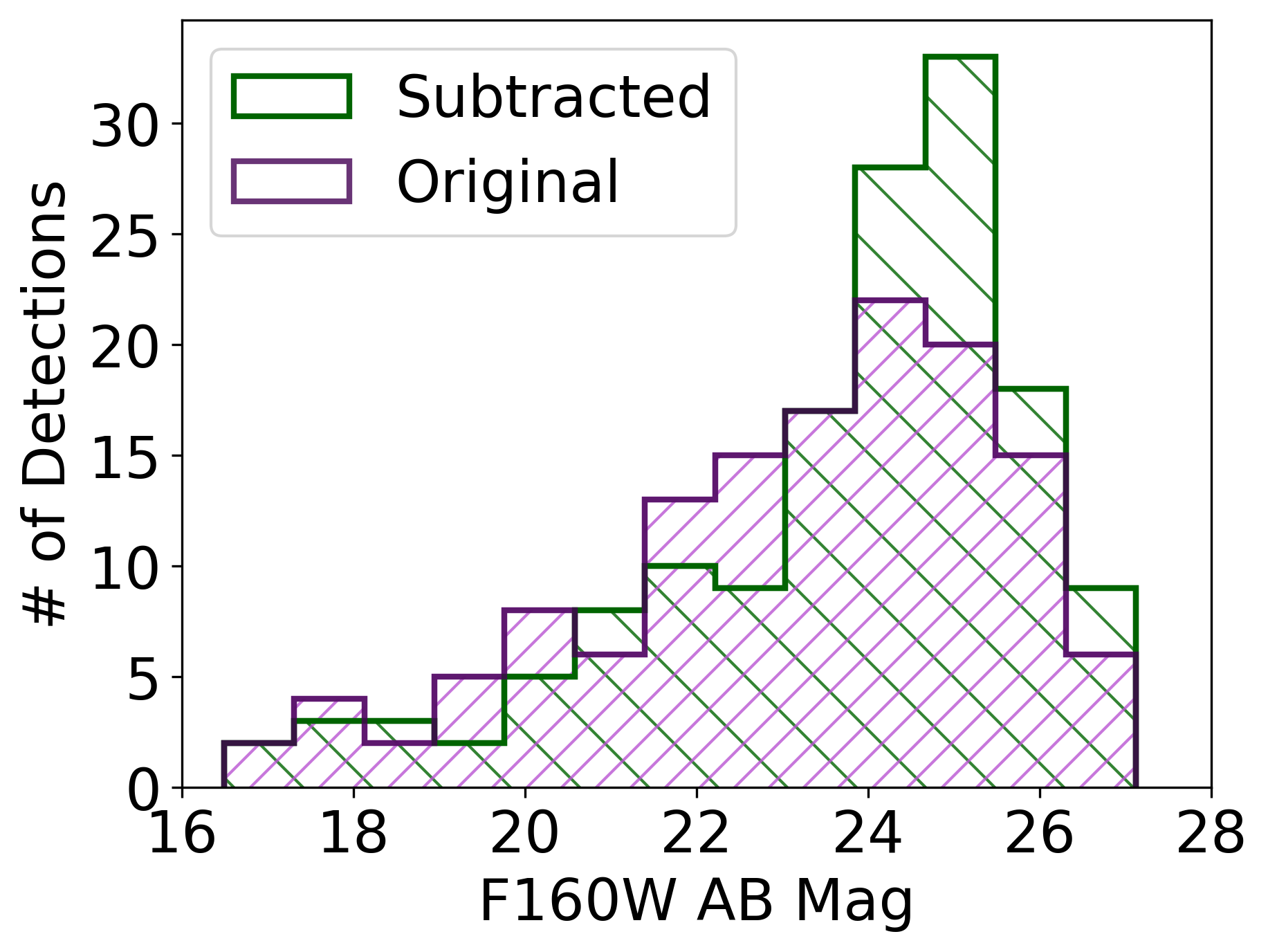}
	\caption{Object counts in the central cluster region as a function
of $F160W$ AB mag (purple) relative to object counts extracted from the
background-subtracted image (green). We delineate the central region by a
72$^{\prime\prime}$\,$\times$\, 24$^{\prime\prime}$ rectangle that runs
along the long-axis of the cluster, covering the central LRGs on both the
NE and SW sides of the cluster (Panel~A of Figure~\ref{fig:imprep}). We
detect 147 objects in the processed image, equating to a 10\% increase from
the 135 objects detected in the original image using the same SExtractor
parameters. We measure more accurate photometry of central objects by their
reduced contamination from the background light, and detect fainter objects,
thereby pushing detections to fainter limiting magnitudes. We refer to
Figure~\ref{fig:objcomp} for specific examples.}
	\label{fig:objcount}
\end{figure}

\begin{figure}[h]
	\centering\includegraphics[scale =1.1]{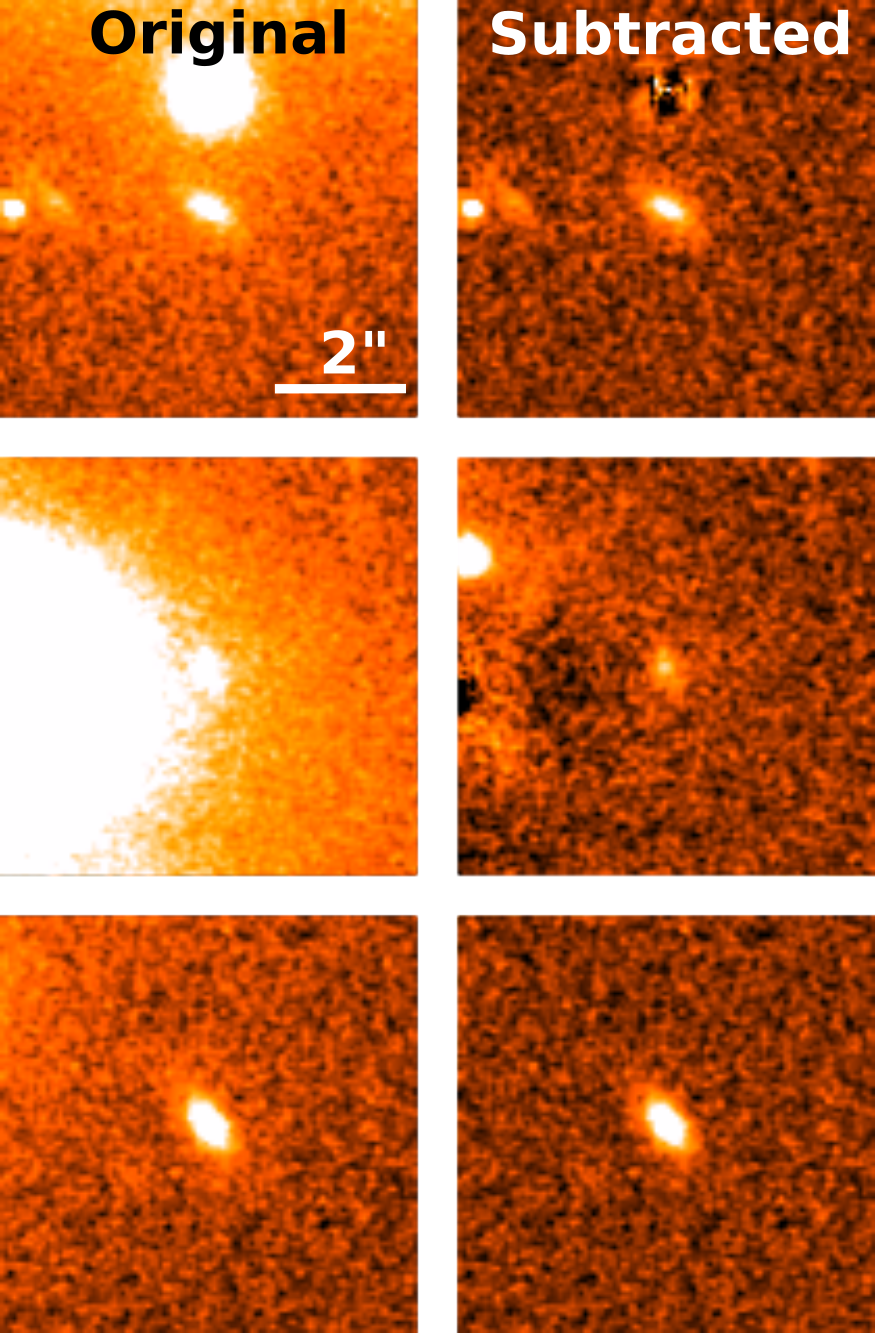}
	\caption{Images centered on three of the 964 galaxies detected in
the HRI ({\it HST} WFC3-IR $F160W$) in the initial image (left-hand side,
drawn from Panel A in Figure~\ref{fig:imprep}), and upon subtraction of the
background (right-hand side, drawn from Panel D in Figure~\ref{fig:imprep}).
In each example, the photometry of the background-subtracted image is
fainter by at least 0.5 AB mag as a result of removing the contaminating
light. A second effect of the background-subtraction is to enable detection
of fainter sources because the noise level decreases.  Both effects operate
in the direction of increasing number counts in the fainter magnitude bins
(Figure \ref{fig:objcount}).
}
	\label{fig:objcomp}
\end{figure}

The two catalogs and their SExtractor segmentation maps are combined
non-redundantly following the methods of \texttt{Galapagos}. The
\texttt{COLD} mode sources are all catalogued, and only \texttt{HOT} mode
sources outside the Kron ellipses of the \texttt{COLD} run are accepted. The
combined catalog contained 1228 sources. We clean this list by hand to
remove artifacts such as edge effects, diffraction spikes, and residuals
from the BCG subtraction. The final detection catalog contains 964+13
sources, with an average aperture Kron ellipse radius of
0.2$^{\prime\prime}$. The `+13' here refers to the initial 13 bright sources
in the image which are separately modeled by \texttt{Galfit}, and are
subtracted prior to detection with the \texttt{HOT+COLD} method. We find the
bright galaxy light and ICL subtraction described in \S3 yields larger
numbers of faint objects in the central region of the cluster

as is demonstrated in Figure
\ref{fig:objcount}. The photometry of faint objects is also more accurately
recovered, 

as depicted in
Figure~\ref{fig:objcomp}. This HRI photometric catalog and its accompanying
segmentation map serve as the inputs for the multi-band photometry in
\texttt{T-PHOT}. However, the default SExtractor segmentation map can cause
inaccuracies in \texttt{T-PHOT} photometry for smaller sources
\citep{Galametz2013}. To complete its setup, we had to modify the
segmentation map of the HRI slightly by scaling the source area in the map
with the \texttt{dilate} software \citep{DeSantis2007}.

\vspace{5mm}
\subsection{Image Alignment and PSF Constructions}
PSFs need to be constructed for each LRI to prepare the images for
simultaneous photometry.  We start by aligning the images in each filter to
the HRI in a two step process using the astropy package \texttt{reproject}
and \texttt{astroalign} tasks.  We find \texttt{astroalign} outperforms
\texttt{reproject} in producing astrometrically precise image alignment, but
fails when images are significantly misaligned.  Hence, we first apply
\texttt{reproject}, which transforms images to the same orientation and
pixel scale based on the WCS header information.  We then follow up on this
step by applying the \texttt{astroalign} task in order to triangulate the
source centroid positions and to calculate a revised value for the
reprojection according to \citet{Beroiz2020}.  We describe below the process
of obtaining a good sampling of the PSF for each of the seven bands.

For the LBT/LBC $gi$ filters, we extract an initial catalog of stars using
the \texttt{DAOStarFinder} subroutine according to \citet{Stetson1987}. We
then visually inspect the output star list and cull out by hand a set of 100
suitable (well isolated and unsaturated) stars.  Cutouts centered on each
star are extracted and are subsequently median-subtracted and normalized to
lessen the impact from image contamination. Finally, the stellar cutouts are
stacked into a single oversampled PSF using the \texttt{EPSFBuilder}
function from the \texttt{photutils} python package.

For {\it HST} WFC3-IR $F110W$ and $F160W$, we adopt the PSFs provided by
STScI\footnote{Empirical Models for the WFC3/IR PSF J.~Anderson 08 Aug
2016}. These nine PSFs sample the profiles in different locations across the
detector.  We average all nine of them together to produce a single
representative PSF.  One averaged PSF is taken for each exposure that makes
up the coadded image, and the PSF is rotated as needed, based on the WCS
image header information.  The individual exposure PSFs are subsequently
averaged together and weighted by the exposure time to yield the PSF for the
coadded image.

For the LUCI+ARGOS $K$-band image, the 4$^{\prime}$\,$\times$\,4$^{\prime}$
FOV lacked well-isolated, unsaturated stars, and only contained a handful of
stars in total. While we found the \texttt{astropy} stacking method to
produce a reasonable-looking PSF by-eye, we obtained a PSF that produced
flatter residuals in the \texttt{T-PHOT} stage by fitting a 2D Gaussian to
crowded but unsaturated stars in the field.  The resulting Gaussian has a
FWHM of $0.29^{\prime \prime}$, consistent with previous results (F19).

For the {\it Spitzer} IRAC image PSFs, we took two different approaches to
build the PSF. First, we stacked the PSFs of isolated stars similar to the
approach outlined above for the LBT/LBC $gi$ images.  Second, we tried using
a set of 25 analytic PSFs supplied by the Spitzer Science
Team\footnote{irsa.ipac.caltech.edu/data/SPITZER/docs/irac/calibrationfiles\newline
/psfprf/}, which are each rendered onto a 5\,$\times$\,5 pixel grid across
the detector. For each individual exposure that makes up the IRAC coadded
image, we interpolated over this PSF suite to generate the one that most
closely corresponds to the location of each object in our catalog. Each PSF
is then 
rotated according the orientation of its exposure relative to the
coadded image. All of the rotated PSFs are then averaged together, and
weighted by the exposure time. The product is an individual analytical PSF
for each object in the catalog. We ultimately adopt the stacking method
approach for obtaining the PSFs, as it produced noticeably flatter residuals
in the \texttt{T-PHOT} stage. At the same time we acknowledge that the
analytical model method has been demonstrated as preferable in other studies
\citep{Merlin2016a,Merlin2021,Pagul2021}.

\subsection{Photometry starting with the $F160W$ and $F110W$ filters}

The $F110W$ image was prepared by the same process as described in \S3 for
the HRI.  We start by applying the final \texttt{Galfit} models for $F160W$
as the initial models for $F110W$ and fit for all objects simultaneously,
including the ICL.  The resulting models are subtracted from the $F110W$
image and then undergo the same median subtraction process applied to the
$F160W$ band.  $F110W$ is then convolved with a matching kernel to the PSF
of $F160W$.  The matching kernel, $K$, is created by taking the ratio of the
PSF of each image as per the convolution theorem, which states that the
integral of two quantities is equivalent to their multiplication in Fourier
space. Hence, we can write
$$\text{PSF}_{2} = K\bigotimes\text{PSF}_{1}$$
where PSF$_1$ is the $F110W$ PSF and PSF$_2$ is the $F160W$ PSF.  
By the convolution theorem
$$\mathcal{F}(\text{PSF}_{2}) = \mathcal{F}(K) * \mathcal{F}(\text{PSF}_{1})$$
$$K = \mathcal{F}^{-1}\left(\frac{\mathcal{F}(\text{PSF}_{2})}{\mathcal{F}(\text{PSF}_{1})}\right)$$
as desired.  We apply a simple Cosine window function prior to convolution
to eliminate any spurious modes picked up by noise. Generating a new RMS map
for $F110W$ following the convolution is not straightforward, and we choose
to approximate the new map with the original RMS map due to the similarity
in PSF FWHM between $F110W$ and $F160W$.  We then enforce the same apertures
as used on the HRI via SExtractor's dual image mode, selecting the $F160W$
image and its associated weight map for detection but the $F110W$ image and
its weight map for measurements. This forced object catalog makes image
registration more straightforward and alleviates some of the issues relating
to image blending, yielding more accurate measured fluxes across both
images. We cite the $F160W$ magnitude as the AUTO magnitude, which is
measured selecting the Kron-like elliptical aperture, and we calculate
$F110W$ magnitude by computing the difference in isophotal aperture
magnitudes between the $F160W$ and $F110W$ bands added to the $F160W$ AUTO
magnitude. We justify the isophotal selection aperture in this case because
it is better at measuring colors, while the AUTO aperture better measures
the total fluxes \citep{Coe2006}.  Magnitude errors are derived using
SExtractor, with the slightly larger $F110W$ errors resulting from the
propagation of errors of each quantity used.

\begin{figure*}
\centering\includegraphics[scale=0.105]{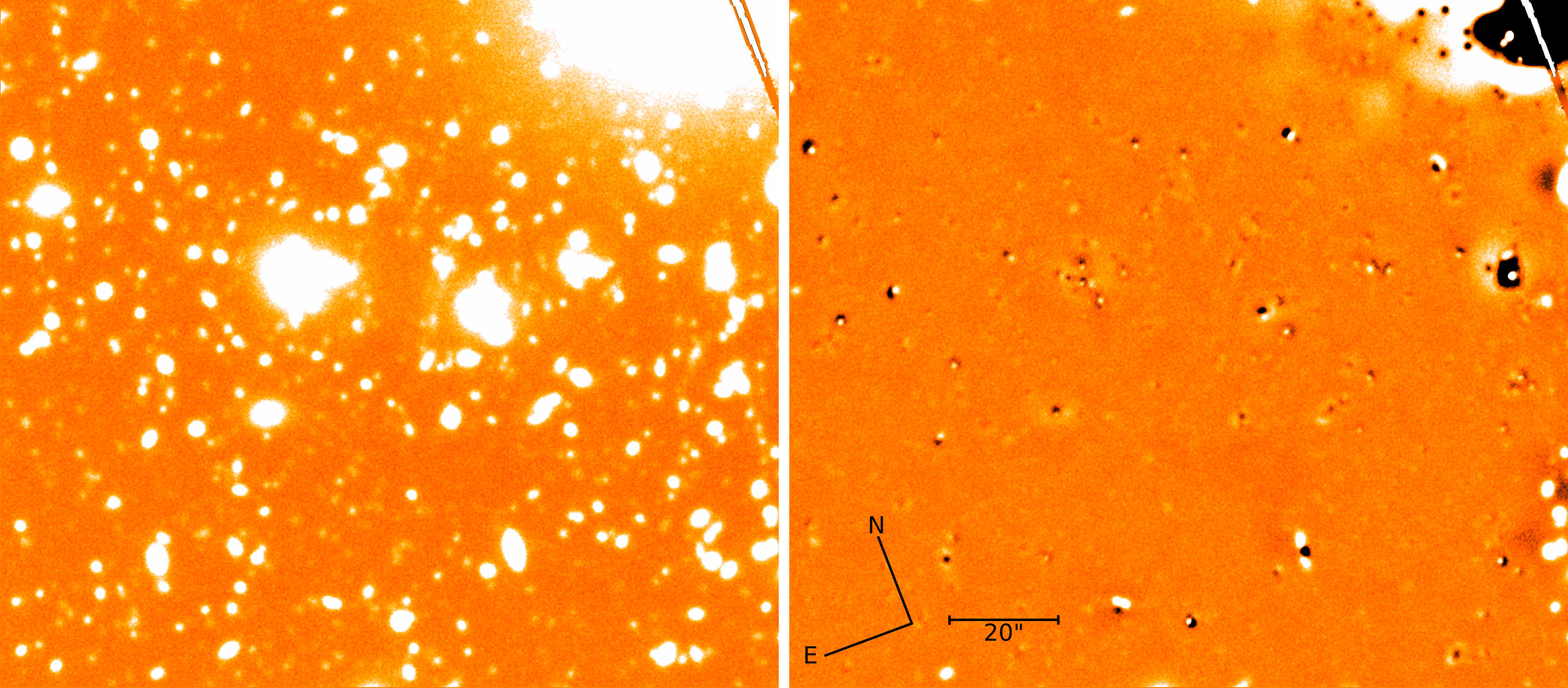}
\caption{Quality check on the \texttt{T-PHOT} photometry for the LBT/LBC
{\it g}-band LRI (left panel). \texttt{T-PHOT} creates a photometric
collage from the HRI ($F160W$) and scales it to minimize the residual
between the collage and the LRI. The residual is flat (right panel),
indicating a good fit of the image priors to the data across the field
(\S4.5).
}
\label{fig:tphot}
\end{figure*}

\subsection{Photometry incorporating the other LRIs}
The LBT/LBC $gi$, LBT/LUCI $K$, and {\it Spitzer} IRAC $Ch1$ and $Ch2$ bands
all have larger PSF sizes relative to the {\it HST} images, making image
blending more of a problem. We perform the multi-band photometry using
\texttt{T-PHOT} in a similar way across all five of these LRIs. The HRI
($F160W$) catalog supplies the positions and fluxes of cutouts for the real
prior, while the HRI \texttt{Galfit} models are applied as the analytical
prior. Both sets of priors are held fixed in all \texttt{T-PHOT} runs across
all LRIs. The only difference between these \texttt{T-PHOT} runs is the
required matching kernel for degrading the HRI cutouts to the LRI PSF, which
we already derived in \S4.3.  For $giK$, the matching kernel is constructed
using the filter's data-derived PSF and the analytical $F160W$ PSF through
the Fourier methods explained in \S4.4. For the {\it Spitzer}/IRAC channels,
the PSF itself is applied as the matching kernel owing to its large profile
compared to $F160W$ \citep{Galametz2013}.
 
We initiated a series of two \texttt{T-PHOT} runs on the 5-filter image
stack. \texttt{T-PHOT} photometry is especially sensitive to the LRI
background, hence the purpose of the first \texttt{T-PHOT} run is to
estimate the background of the LRI using the built-in option for local
background estimation. Flat background object counterparts are assigned to
each object in the input catalog. All objects are fit simultaneously,
returning the 2D image collage as outlined in \S4.1.  We modify this collage
to only retain the fitted flat background objects, yielding the background
model.  This model is subsequently smoothed using a 2D Gaussian kernel to
produce a relatively smooth approximation of the image background. This
background model is subtracted from the image to yield the final
background-subtracted image.

To produce the multi-band photometry, we perform a second \texttt{T-PHOT}
run on the background-subtracted images of the 5-filter suite. As a gauge of
the quality of this run, the original LRI is compared to the LRI with the
\texttt{T-PHOT} collage subtracted from it. A flat residual image, such as
the one shown in Figure~\ref{fig:tphot} for the $g$-band, indicates a good
fit of the image priors to the data across the field. This procedure allows
for iterative modifications to the image priors until a satisfactory
residual is obtained.

Each \texttt{T-PHOT} run returns a covariance index that correlates with the
quality of the fit. The covariance index is sensitive to the effects of
crowding on the \texttt{T-PHOT} fit, such that covariance indices greater
than unity suggest a high degree of degeneracy between the fit photometry of
neighboring objects.  We cite photometric errors according to the output
from \texttt{T-PHOT}. We stress that these uncertainties are statistical
relative to the input RMS map, and do not include systematic errors, such as
artifacts introduced by the PSF estimation, or any remaining non-uniform
background in the $F160W$ image cutouts.  Systematics are broadly accounted
for in the photometric redshift estimation by enforcing a floor minimum
uncertainty level for the input photometric measurements amounting to 0.05
mag for the $F110W$, $F160W$, and $K$ filters, and 0.1 mag for the $gi$ and
IRAC filters as in \cite{Merlin2016a}, \cite{Bradac2019} and
\cite{Pagul2021}.

\subsection{Preparation for Photometric Redshift Estimation} \label{sec:zphot}
We estimate photometric redshifts with the Bayesian Photometric Redshift
(BPZ) software package \citep{Benitez2000, Benitez2004,Coe2006}. BPZ fits
photometric measurements to SEDs of different galaxy types using Bayesian
inference. Crucially, the Bayesian approach utilizes assumed priors on
$i$-band brightness and galaxy type such that only the SEDs which best
represent the input photometry are fit to the data. BPZ produces a best fit
redshift and a galaxy type by interpolating over SED templates provided by
\cite{Coleman1980}, \cite{Kinney1996}, and \cite{Bruzual2003}. In
preparation to doing accurate photometric redshift estimation with BPZ, we
also make adjustments to the photometric zeropoints and apertures.

\subsubsection{Zeropoint Correction}
We initially trained BPZ on a subset of 20 objects which also have
spectroscopic redshifts.  This involved fixing the object redshift to the
spectroscopic redshift and then enforcing BPZ to fit the SEDs by only
allowing the galaxy type to vary.  In turn the BPZ code outputs suggestions
of any zeropoint magnitude offsets.  This is an iterative process in which
an offset is computed, and then the code is rerun, until the suggested
offset approaches zero.

We found the magnitude offsets to be small or zero in most bands. We did,
however, notice non-negligible suggested offsets of $\sim0.35$ and
$\sim0.25$ magnitudes in LBT $g_{AB}$ and $i_{AB}$, respectively. To explore
this difference, a second test was conducted comparing the photometry in the
central the central 5\,$\times$\,5 arcmin FOV with SDSS.  We refer to
\S2.1.1 for details. Somewhat reassuringly, we obtain a similar zeropoint
offset by both approaches. We opted to apply the offsets measured via the
SDSS comparison for our photometry. The revised values for the zeropoint
magnitudes are reported in Table \ref{tab:1}.

Our BPZ calibration also motivated small zeropoint corrections in the IRAC
images of 0.21 and 0.14 for Ch\,1 and Ch\,2, respectively. On further
inspection, we confirmed that residuals from \texttt{T-PHOT} were generally
more negative than the sky in those bands. We believe this is due to the
unique structure and large width of the IRAC PSF.  At the same time,
systematic offsets in magnitude are not uncommon. For example, in one study
of the HFF clusters, a minor IRAC offset was also reported of a similar
degree but in the opposite direction, potentially owing to their
analytical-style approach to estimate the PSF rather than our data-driven
one as described in \S4.3 \citep{Pagul2021}.

On applying all the zeropoint corrections to the data, the BPZ code is
re-run with the full photometric catalog, freely and without any fixed
redshifts, to create the photometric redshift catalog.  In doing so, we
impose a few criteria to ensure a valid measurement of the photometry.
First, the photometry of objects whose ratios of flux to flux error are less
than the measured image signal-to-noise ratio are classified as
non-detections and are removed.  We also throw out candidate detections for
which the flux error is negative or unreasonably large, or the covariance
index output by \texttt{T-PHOT} is larger than unity.

\subsubsection{Custom Apertures on Arcs}

The extraction apertures were tailored for the three sets of image
multiplicities for which we achieve quality photometric redshift estimation:
Arcs 1a/b, Arcs 11a,b, and Arcs 2a,c (Panel D of
Figure~\ref{fig:imprep}). The infrared images of Arc 1a/b consist of two
blue knots and a redder and more diffuse underlying component with a total
angular extent of 5 arcseconds.  Because the \texttt{T-PHOT} fitting method
integrates across the image, the multi-component information gets
subsequently lost. 
To recover the input
redshift we found it necessary to place down one aperture for each knot as
well one for the overall arc-shape.  We refer to \S4.6.2 below for details.

\begin{figure}[h]
	\centering\includegraphics[scale =.555]{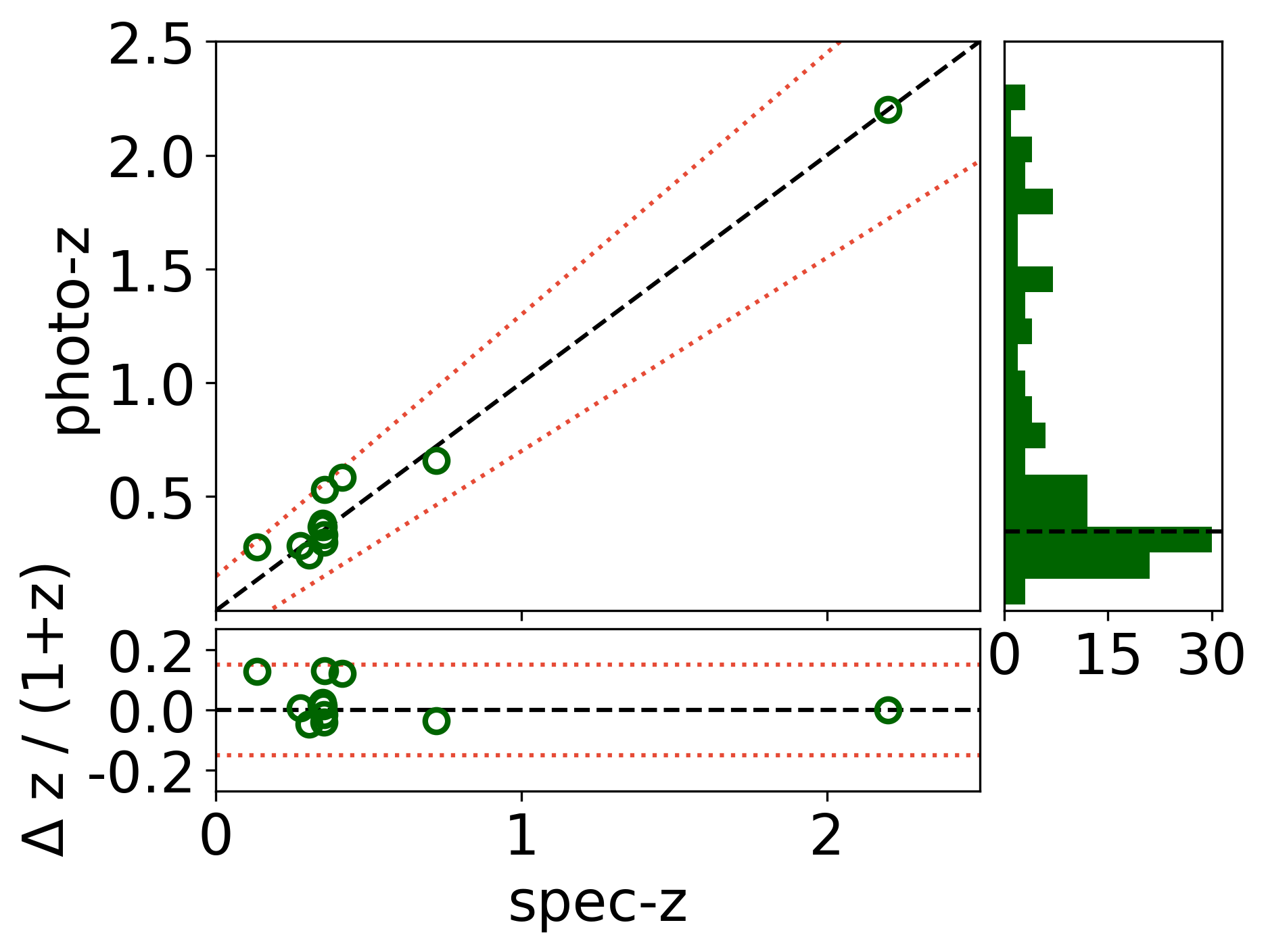}
	\caption{
	Comparison of the training set of galaxy redshifts with their
photometric redshift estimates obtained in this study, up to $z$\,=\,2.5.
We define a metric for the goodness-of-fit as
$\vert\delta z\vert/(1+z)$\,$<$\,0.15  (dashed lines) to assess the overall
performance of the photometry. The histogram on the right-hand-side depicts
the distribution of photometric redshifts, which peaks at the cluster
redshift  $z$\,=\,0.348 (black dashed line).  We find the photometric
redshift estimations to be contained within the goodness-of-fit metric for
all sources.}
	\label{fig:zcomp}
\end{figure}

\begin{figure}[h]
	\centering\includegraphics[scale =0.6]{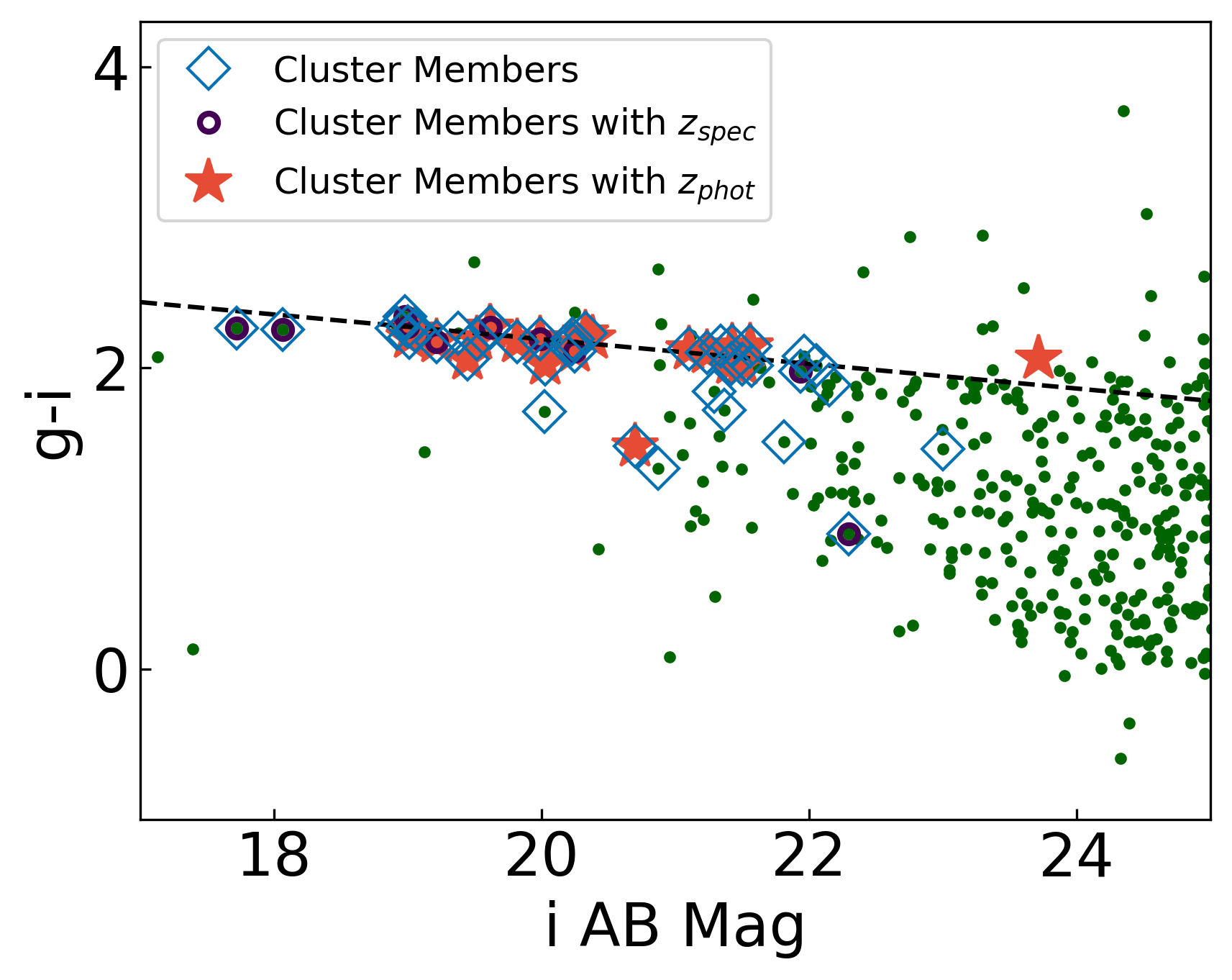}
	\caption{LBT/LBC {\it g-i} colors for \texttt{T-PHOT} photometry
performed on objects within the {\it HST} FOV (green filled circles). The
spectroscopically confirmed cluster members are depicted by the purple open
circles, while the photometrically confirmed cluster members are indicated
by the red stars. The cluster red sequence  is depicted for reference
(black dashed line), which guides the selection
of cluster members {applied to our lens model}
(blue diamonds).}
	\label{fig:cmd}
\end{figure}

\begin{figure*}[h]
	\centering\includegraphics[scale =0.56]{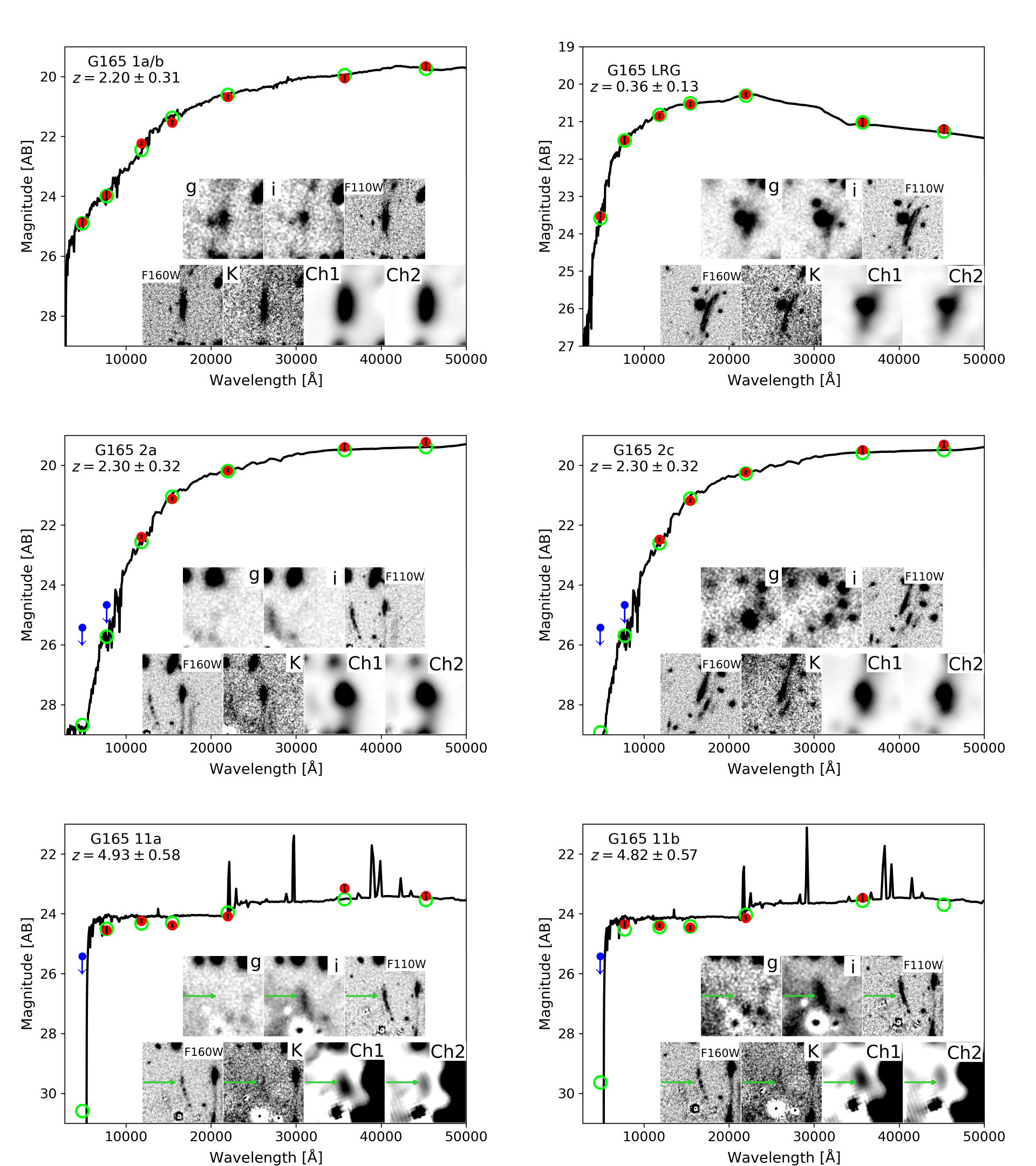}
	\caption{Best fit BPZ SEDs from SExtractor and \texttt{T-PHOT}
photometry of Arcs 1a/b, 2a, 2c, 11a, and 11b and a typical example of a
cluster LRG. The best fit SEDs (black line) interpolated from
\cite{Kinney1996}, \cite{Bruzual2003}, and \cite{Coleman1980} templates
satisfy our criterion for secure redshifts by returning a reduced
$\chi^2$\,$<$\,1 while using a minimum of 5 bands. SED fit quality is
determined by the offset between the photometry estimated for the SED in
a given filter (green circles) and the photometry measured through
SExtractor and \texttt{T-PHOT} (red dots), as well as by satisfying the BPZ
prior on the $i$-band magnitude. Drop outs (blue arrows) are reported where
the measured magnitude falls below the measured 3-$\sigma$ detection
limit. Image stamps of each object are shown in the insets with the filter
as labeled. Photometric redshifts are obtained for these five members of
three different image multiplicities, which contribute tighter constraints
to the lens model.}
	\label{fig:sed}
\end{figure*}

On other image multiplicities, Arcs 11a,b consists of two bluer knots on top
of a more extended, and somewhat redder, stellar continuum. For this case,
we
lay down four apertures, one for each of the knots. We choose not to put
down an aperture over the larger arc structure because the surface
brightness is low and in principle a photometric redshift can be obtained
from any of the knots. This choice of a small aperture is also advantageous
for minimizing background contamination.  Arcs 2a 
and 2c are compact
and bright, making them straightforward to detect using SExtractor. Given
the compact morphology, further adjustments to the apertures or photometry
were unnecessary.
\vspace{12mm}
\section{{Photometric Results}} \label{sec:phot_results}
\subsection{Photometric Catalogs}
We estimate photometric redshifts for 143 galaxies, for which each galaxy is
detected in at least five bands and has been confirmed to be a real source
by our visual inspection. Furthermore, each SED fitting result must satisfy
the condition that the modified $\chi^2<1$, a conservative cutoff to ensure
high quality fits with BPZ introduced in \cite{Coe2006}. The distribution of
photometric redshifts peaks at the cluster redshift, and has a long tail
extending to higher values (Figure~\ref{fig:zcomp}).  Of these, 22 objects
are found to be in or near the cluster redshift, with 37 in the foreground
and 84 in the background. On background objects, 42 objects are at
1\,$<$\,$z$\,$<$\,3, with four high redshift sources at $z$\,$>$\,5. The
result of this study is that new optical-IR photometric redshifts are
obtained for four lensed galaxy images from two image systems for which
there was no previous photometric or spectroscopic values. These new values
contribute tighter constraints to the lens model (\S5.3, \S7).

\begin{deluxetable}{ccccccccccc}[h]
\rotate
\tablecaption{High-z Photometry From G165 Field}
\tablecolumns{11}
\tablehead{
\colhead{ID}$^a$ & \colhead{R.\,A.} & \colhead{Decl.}  & \colhead{$z_{phot}^b$} & \colhead{$g_{AB}^c$} & \colhead{$i_{AB}^c$}& \colhead{$F110W_{AB}$} & \colhead{$F160W_{AB}$} & \colhead{$K_{AB}$} & \colhead{$Ch1_{AB}$} & \colhead{$Ch2_{AB}$}
}
\startdata
$1^d$ & 171.8113 & 42.47299 & $2.20^{+0.31}_{-0.31}$ & $23.96\pm0.11$ & $24.86\pm0.08$ & $22.23\pm0.02$ & $21.54\pm0.05$ & $20.68\pm0.05$ & $20.06\pm0.01$ & $19.65\pm0.01$ \\
$2^e$ & 171.81645 & 42.47473 & $2.30^{+0.32}_{-0.32}$ & $>26.1$ & $>26.8$ & $22.39\pm0.00$ & $21.12\pm0.00$ & $20.18\pm0.00$ & $19.39\pm0.00$ & $19.22\pm0.00$ \\
$3^e$ & 171.81384 & 42.47811 & $2.30^{+0.32}_{-0.32}$ & $>26.1$ & $>26.8$ & $22.48\pm0.00$ & $21.18\pm0.00$ & $20.24\pm0.00$ & $19.49\pm0.00$ & $19.31\pm0.00$ \\
$4^f$ & 171.81567 & 42.47835 & $4.93^{+0.58}_{-0.58}$ & $24.55\pm0.01$ & $>26.8$ & $24.24\pm0.01$ & $24.38\pm0.03$ & $24.09\pm0.06$ & $23.15\pm0.04$ & $23.40\pm0.03$ \\
$5^f$ & 171.81572 & 42.47802 & $4.82^{+0.57}_{-0.57}$ & $24.35\pm0.01$ & $>26.8$ & $24.40\pm0.01$ & $24.46\pm0.03$ & $24.14\pm0.07$ & $23.46\pm0.05$ & --- \\
6 & 171.81575 & 42.47784 & $5.06^{+0.59}_{-0.59}$ & $25.11\pm0.02$ & $>26.8$ & $24.94\pm0.01$ & $24.87\pm0.04$ & $24.61\pm0.10$ & $24.73\pm0.10$ & --- \\
7 & 171.8211 & 42.45727 & $1.86^{+0.29}_{-0.28}$ & $25.67\pm0.02$ & $26.34\pm0.02$ & $25.28\pm0.02$ & $24.71\pm0.03$ & $24.69\pm0.07$ & $24.42\pm0.04$ & $24.21\pm0.03$ \\
8 & 171.84108 & 42.46482 & $1.81^{+0.28}_{-0.55}$ & $24.70\pm0.01$ & $24.98\pm0.01$ & $24.13\pm0.01$ & $23.76\pm0.02$ & $23.62\pm0.03$ & $23.46\pm0.02$ & $23.49\pm0.01$ \\
9 & 171.8184 & 42.45991 & $1.99^{+0.48}_{-0.35}$ & $27.54\pm0.12$ & $27.47\pm0.05$ & $26.88\pm0.04$ & $26.42\pm0.10$ & --- & $25.98\pm0.14$ & --- \\
10 & 171.83697 & 42.46778 & $4.82^{+0.57}_{-0.57}$ & $26.07\pm0.03$ & $>26.8$ & $25.94\pm0.02$ & $25.78\pm0.06$ & $25.67\pm0.15$ & $24.79\pm0.07$ & $24.83\pm0.06$ \\
11 & 171.80296 & 42.4641 & $2.00^{+0.29}_{-0.29}$ & $27.30\pm0.11$ & $>26.8$ & $24.04\pm0.01$ & $22.79\pm0.01$ & $22.12\pm0.01$ & $21.45\pm0.00$ & $21.34\pm0.00$ \\
12 & 171.81299 & 42.471 & $2.05^{+0.30}_{-0.30}$ & $23.91\pm0.01$ & $24.28\pm0.00$ & $23.40\pm0.01$ & $22.94\pm0.01$ & $22.66\pm0.01$ & $22.54\pm0.01$ & $22.52\pm0.01$ \\
13 & 171.81502 & 42.47611 & $2.06^{+0.30}_{-0.30}$ & $25.02\pm0.02$ & $26.60\pm0.03$ & $22.05\pm0.00$ & $20.82\pm0.00$ & $19.86\pm0.00$ & $19.13\pm0.00$ & $19.00\pm0.00$ \\
14 & 171.79628 & 42.46708 & $1.94^{+0.29}_{-0.29}$ & $23.90\pm0.01$ & $24.69\pm0.00$ & $23.67\pm0.01$ & $23.10\pm0.01$ & $23.01\pm0.02$ & $22.60\pm0.01$ & $22.57\pm0.01$ \\
15 & 171.8178 & 42.47324 & $1.81^{+0.27}_{-0.27}$ & $26.01\pm0.03$ & $28.87\pm0.18$ & $23.14\pm0.01$ & $22.26\pm0.01$ & $21.71\pm0.01$ & $21.26\pm0.00$ & $21.19\pm0.00$ \\
16 & 171.83023 & 42.47862 & $1.80^{+0.27}_{-0.27}$ & $23.56\pm0.00$ & $23.81\pm0.00$ & $23.27\pm0.01$ & $22.95\pm0.01$ & $22.89\pm0.02$ & $22.58\pm0.01$ & $22.61\pm0.01$ \\
17 & 171.79925 & 42.47057 & $2.15^{+0.31}_{-0.31}$ & $25.28\pm0.02$ & $26.19\pm0.02$ & $23.80\pm0.01$ & $22.59\pm0.01$ & $22.00\pm0.01$ & $21.29\pm0.00$ & $21.08\pm0.00$ \\
18 & 171.81919 & 42.4768 & $3.12^{+0.40}_{-0.40}$ & $24.97\pm0.02$ & $26.13\pm0.03$ & $25.21\pm0.03$ & $24.44\pm0.05$ & $23.54\pm0.17$ & --- & $22.96\pm0.01$ \\
19 & 171.82957 & 42.48002 & $1.82^{+0.28}_{-0.28}$ & $26.20\pm0.04$ & $26.59\pm0.03$ & $24.57\pm0.01$ & $23.74\pm0.02$ & $23.22\pm0.02$ & $22.77\pm0.02$ & $22.53\pm0.01$ \\
20 & 171.81498 & 42.47713 & $3.85^{+0.48}_{-3.36}$ & $24.52\pm0.01$ & $26.09\pm0.02$ & $24.67\pm0.01$ & $24.64\pm0.03$ & $24.15\pm0.06$ & --- & --- \\
21 & 171.82221 & 42.48107 & $1.95^{+0.29}_{-0.29}$ & $24.43\pm0.01$ & $24.88\pm0.01$ & $23.53\pm0.01$ & $22.85\pm0.01$ & $22.52\pm0.02$ & $22.11\pm0.01$ & $22.16\pm0.01$ \\
22 & 171.79848 & 42.48498 & $5.24^{+0.61}_{-0.61}$ & $26.62\pm0.07$ & --- & $25.90\pm0.02$ & $25.68\pm0.09$ & $25.68\pm0.15$ & $24.60\pm0.10$ & $24.24\pm0.06$ \\
23 & 171.80525 & 42.48604 & $1.85^{+0.28}_{-0.28}$ & $24.78\pm0.01$ & $26.14\pm0.02$ & $22.93\pm0.01$ & $22.03\pm0.01$ & $21.37\pm0.01$ & $20.53\pm0.00$ & $20.41\pm0.00$ \\
24 & 171.82915 & 42.49285 & $3.34^{+0.42}_{-0.91}$ & $25.82\pm0.03$ & $25.98\pm0.01$ & $25.75\pm0.02$ & $25.70\pm0.10$ & --- & $25.46\pm0.09$ & $25.57\pm0.08$ \\
25 & 171.83135 & 42.49005 & $3.37^{+0.43}_{-0.43}$ & $25.89\pm0.03$ & $26.90\pm0.03$ & $25.59\pm0.02$ & $25.39\pm0.06$ & --- & $24.49\pm0.04$ & --- \\
26 & 171.79051 & 42.47874 & $2.79^{+0.40}_{-0.37}$ & $27.03\pm0.12$ & $28.91\pm0.35$ & $27.39\pm0.07$ & $26.01\pm0.08$ & $25.16\pm0.11$ & --- & --- \\
27 & 171.80875 & 42.47923 & $1.81^{+0.28}_{-0.53}$ & $24.29\pm0.01$ & $24.70\pm0.01$ & $23.87\pm0.01$ & $23.44\pm0.02$ & $23.54\pm0.03$ & $23.07\pm0.02$ & $23.12\pm0.01$ \\
28 & 171.78579 & 42.47477 & $5.65^{+0.65}_{-4.25}$ & $28.23\pm0.22$ & --- & $26.61\pm0.02$ & $26.53\pm0.15$ & $26.13\pm0.20$ & $25.21\pm0.07$ & --- \\
\enddata
\tablenotetext{a}{Object ID number.}
\tablenotetext{b}{Photometric redshift estimates are presented for 26 objects
 with $z_{phot}$\,$>$\,1.8 in at least five bands.} 
\tablenotetext{c}{We assign lower limits to be a 3$\sigma$ limiting magnitude}
\tablenotetext{d}{This is the DSFG Arc 1a/b, which has a
$z_{spec}$\,=2.2357\,$\pm$\,0.0002 \citep{Harrington2016} and
$z$\,=\,$2.2362$\,$\pm$ \,0.0001 \citep{Nesvadba2019}; the best fit SED can
be found in Figure~\ref{fig:sed}}
\tablenotetext{e}{Arc 2a and 2b; the best fit SED can be found in Figure
\ref{fig:sed}}
\tablenotetext{f}{Arc 11a and 11b; the best fit SED can be found in Figure
\ref{fig:sed}}
\label{tab:highz}
\end{deluxetable}

\subsection{Selection of Cluster Members}
The input catalog of cluster members for the lens model contains 40
galaxies, which results from a strict color cut following the red sequence
(Figure~\ref{fig:cmd}). We use the 22 photometrically- and 10
spectroscopically-confirmed cluster members (some members have both redshift
types) to guide the color cut.  We find that the cluster member list does
not change drastically from F19, yet still qualifies as an improvement
because it includes 10 previously overlooked cluster members, and removes
two $z \approx 0.48$ interlopers which share similar colors to the cluster
members. Nearly all previously excluded cluster members were relatively
faint, and also failed to meet the more stringent color criteria in F19.
The $g-i$ colors of the spectroscopically confirmed cluster members as well
as the photometrically confirmed cluster members are depicted in Figure
\ref{fig:cmd}.

\subsection{The Image Multiplicities}
We set out to estimate photometric redshifts for all the images of all the
image multiplicities, and succeeded in obtaining redshift values for Arc
1a/b, Arc 2a, Arc 2c, Arc 11a, and Arc 11b, which for reference are labeled
on Figure~\ref{fig:imprep} in Panel D, and are depicted as image stamps in
Figure~\ref{fig:sed}.  Of these, Arc 1a/b is the only member of any image
system with a spectroscopic redshift \citep{Harrington2016, Harrington2021}.
Its magnification factor is estimated from our lens model to be
$\mu$\,$\approx25$.  The morphology of Arc 1a/b consists of two bluer,
star-forming knots opposite the critical line.  {As a sanity check on the
photomety, we estimate a photometric redshift for Arc 1a/b} of
$z$\,=\,$2.20\,\pm0.31$, which is consistent with its spectroscopic
redshift.  The counterimage, Arc 1c, is detected in $K$-band and $Ch1[3.6]$
and $Ch2[4.5]$ data, and has the similar colors and model-predicted location
expected of a counterimage. However it is too faint and blended in the bluer
bands to estimate its photometric redshift. The spectroscopic redshift is
preferred for constraining lens model due to its much lower uncertainties.

Arcs 2a, 2b, and 2c belong to a single image system. They have a mean
$K_{AB}$ magnitude of $\approx$\, 20.2, and high magnification factors of
$\mu$\,$\approx10$.  SED fits are made for Arcs 2a and 2c, for which we
estimate photometric redshifts of $z$\,=\,2.30\,$\pm$\,0.32 for both images
(Figure~\ref{fig:sed}).  The similarity in redshift between Arcs 2a,c is
expected under the interpretation that they are multiple images of the same
background galaxy.  According to the morphology, colors, and the lens model
predictions, Arc 2b is also a member of this image system, but an SED fit
was not made for this image family member owing to obscuration by a bright
cluster galaxy.

Arcs 11a and 11b have a similar morphology consisting of a pair of compact
knots that is doubly-imaged about the critical curve. These galaxy images
are somewhat bluer and are more or less flat in $F110W-F160W$, suggesting
that the source is a star forming galaxy that is relatively unobscured by
dust. We measure high magnification factors of $\approx50$ and $\approx60$
for Arcs 11a, and 11b, respectively. Figure \ref{fig:sed} shows the SED
fits, which return estimates of $z_{phot}$\,=\,$4.93\pm0.58$ for Arc 11a,
and $z_{phot}$\,=\,$4.82\pm0.57$ for Arc 11b. We choose to include $K$-band
photometry for this arc, although it is at or slightly below the detection
limit. 11c is also a member of this image system as predicted by the lens
model, however a quality photometric redshift was not measured for this
system, perhaps owing to its especially low surface brightness compared to
its counterimages.

\subsection{The High-z Arcs}

We estimate photometric redshifts for 28 lensed sources with $z$\,$>$\,1.5,
including 6 lensed sources with $z$\,$>$\,4. Table \ref{tab:highz} gives the
catalog, where the columns are: object name, coordinate, $z_{phot}$, and the
AB magnitude in each of the seven bands. We identify many galaxies at
similar redshifts, yet upon comparing the colors and the morphologies with
the lens model, we do not confirm any new image multiplicities. The number
count at $z$\,$>$\,4 is consistent to order of magnitude with the expected
number for its redshift given by lensing the \cite{Mason2015} luminosity
function with our lens model. We refer to \S7.3 for a more complete
discussion of the predicted galaxy number counts.

\subsection{Comparison to Spectroscopic Members}
We compare the redshifts of the spectroscopic sample with the photometric
redshift catalog.  This exercise provides an independent check on the
overall quality of the photometric redshift fit.  To make this comparison,
we establish the goodness-of-fit metric
$\mid$$\delta$\,$z$\,$\mid$/(1+$z$)\,$<$\,0.15, {which measures the
fraction of outliers following \cite{Castellano2016} and
\cite{Dahlen2013}. We choose to only compare secure photometric
redshifts, meaning that an object is clearly detected in 5 bands and provides a
modified $\chi^2 < 1$.} We find secure photometric redshifts for 12 of the
22 objects in the spectroscopic sample in the HST FOV, all of which pass our
goodness-of-fit metric (Figure~\ref{fig:zcomp}).

\section{{Spectroscopic Results}} \label{sec:spec_results}
\subsection{The MMT/Binospec Sample}
We report secure redshift identifications of the Binospec spectroscopy, by
which we mean that we require two or more spectroscopic features to be
detected at the $>$2$\sigma$ level relative to the continuum.  In the case
of a single emission line source, we require also the detection of a second
significant feature such as a continuum break.  Typical absorption- and
emission-line features detected in these spectra (depending on the redshift
and object type) are: Fe\,II$\lambda \lambda$2587,2600, Mg\,II$\lambda
\lambda$2796,2803, Mg\,I$\lambda$2852, [O\,II]$\lambda \lambda$3727,3729,
Ca\,H\,\&\,K, G-band, [O\,III]$\lambda\lambda$4959,5007,
Mg\,I$\lambda\lambda\lambda$5167,5173,5184, NaD, O\,I$\lambda$6300,
[N\,II]$\lambda\lambda$6548,6583, Balmer family (H$\alpha$ through
H$\theta$), and [S\,II]$\lambda\lambda$6716,6731.

The analysis yielded 86 redshifts.  These redshifts separate out into 17
galaxies in the redshift range of the cluster of 0.330\,$<$\,$z$\,$<$\,0.366
at any {radius from the cluster center (clustercentric radius)}, and 51
sources with 0.366\,$<$\,$z$\,$<$\,1.13. Nineteen objects have redshifts
which place them in the foreground of the lens.  The Dusty Star Forming
Galaxy (DSFG) Arc 1a was targeted, but did not yield a redshift.  No
spectroscopic features are detected in its spectrum over the wide wavelength
baseline of 3900\,--\,9200 \AA. This result is not surprising given that the
expected position of Ly$\alpha$ falls just blueward of the spectral
bandpass, and the nebular [O\,II] line is redshifted out of the bandpass.

\begin{figure}[h]
	\centering\includegraphics[scale =0.63]{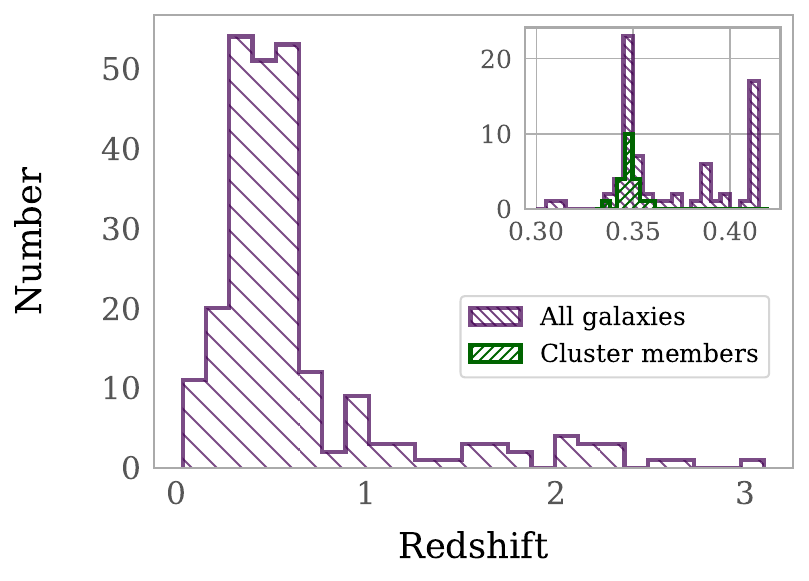}
	\caption{Spectroscopic redshift histogram for galaxies up to
$z$\,=\,3 for the 273 redshifts drawn from all available sources.  A peak
appears at the cluster redshift of 0.348. A total of 39 galaxies meet the
velocity criterion for cluster membership, of which only half meet also the
{radial cut} (green-hashed histogram in the inset plot).  We
note that a smaller peak is identified at the slightly higher redshift of
$z$\,=\,0.41. On further investigation, the vast majority of these galaxies
are situated at large clustercentric radii of $\apg$1~Mpc.  These galaxies
may be part of a larger scale structure, but are not expected to contribute
a significant strong lensing effect.}
	\label{fig:zhist}
\end{figure}

\vspace{5mm}

\subsection{The Master Spectroscopic Sample}

We combine the MMT/Binospec redshifts with other redshifts available in the
literature (F19 and SDSS DR-16), which sum up to 273 spectroscopic
redshifts.  Cluster membership is met for galaxies which have velocities
within $\pm$\,4000~km\,s$^{-1}$ about the mean value of $z$\,=\,0.348.  This
corresponds to range of 0.330\,$<$\,$z$\,$<$\,0.366. We choose $\pm$\,4000
km s$^{-1}$ so as to include sources in the outskirts of this cluster with
its somewhat elongated structure.  Of the 273 redshifts, 39 galaxies make
the redshift cut.  We further downselect to the subset of galaxies whose
clustercentric radii are within 1~Mpc. By these criteria, a total of 21
galaxies are admitted to the master cluster member catalog
(Figure~\ref{fig:zhist}, inset). They are comprised of 8 galaxies from
MMT/Binospec (this study), 6 galaxies from Gemini/GMOS (F19), 5 galaxies
from MMT/Hectospec (F19), and 2 galaxies from the SDSS DR 16.

The vast majority of cluster members lie reasonably well on the cluster red
sequence indicated in Figure \ref{fig:cmd}.  The one outlier, a faint
elliptical galaxy with $i_{AB}$\,=\,22.4 mag, is situated in the outskirts
of a luminous galaxy at $z$\,=\,0.033 (SDSS, DR16) that skews its color.
Thirteen cluster members are contained in the FOV of this study, which is
the field of intersection of the 7-band filter suite roughly comparable to
the {\it HST} WFC3-IR FOV.  This number is a factor of three improvement
relative to the number of cluster members reported in the same FOV in F19.
Of the 18 cluster members in their study at any clustercentric radius, three
are removed from consideration in our new member catalog. They are: ``s7"
from MMT/Hectospec, whose spectrum had an insecure redshift, ``s57" from
Gemini/GMOS catalog, and ``s35" from the SDSS catalog which were both made
redundant by our new MMT/Binospec catalog.  There are a set of 22 galaxies
which have spectroscopic redshifts behind the cluster of
0.398\,$<$\,$z$\,$<$\,0.426.  Of these, only one is situated within a
clustercentric radius corresponding to the FOV of this study, and the vast
majority (18 of 22) are situated at clustercentric radii greater than 1~Mpc.
These galaxies are interesting because they may be part of a larger scale
structure, but are not expected to make a significant contribution to the
strong lensing model presented here.

\begin{figure}[h]
	\centering\includegraphics[scale =0.62]{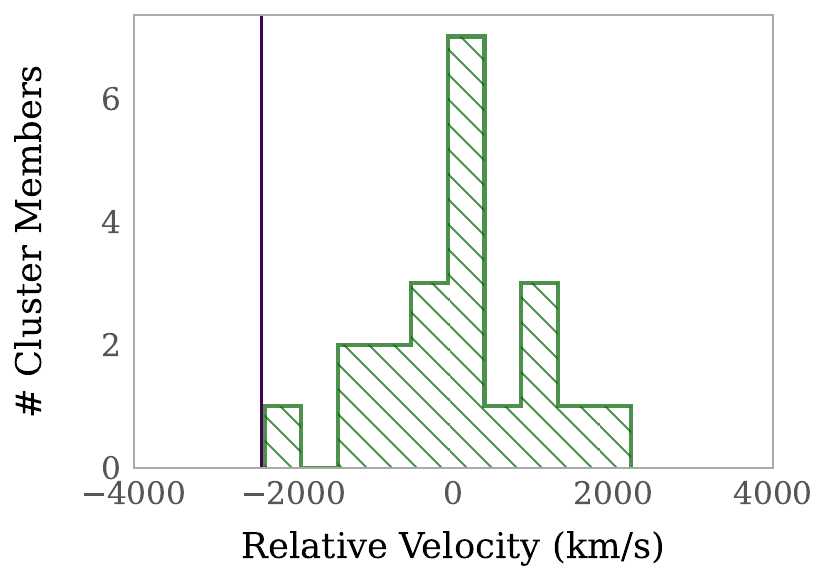}
	\caption{Histogram of the velocities of the confirmed cluster
members in the common field that is covered by all of the data of 600~kpc
(green-hashed histogram of the inset of Figure~\ref{fig:zhist}), plotted
relative to the velocity center.  The BCG is marked by the green vertical
line at -2400 km~s$^{-1}$. The velocities describe a roughly normal
distribution, despite the cluster's visual bimodality.  As such, the large
velocity offset of the BCG suggests a cluster disturbance. Additional clues
as to the cluster's dynamical state are discussed in \S8.}
	\label{fig:vhist}
\end{figure}

The velocities of the cluster members in the common field of interest of
600~kpc are binned and plotted in Figure \ref{fig:vhist}, where for
reference 0~km~s$^{-1}$ marks the cluster's systemic velocity and the solid
vertical line marks the velocity of the brightest cluster galaxy.  Although
G165 is a visual double-cluster, it is a challenge to separate out the two
halves based on their relative velocities. Within radii of 600~kpc and 1~Mpc
we measure velocity dispersions of 1040\,$\pm$\,282 km~s$^{-1}$ and
1010\,$\pm$\,209 km~s$^{-1}$, respectively. We calculate the mass by
applying the virial theorem via the {Gapper method
\citep{Wainer1976,Beers1990}}. We obtain masses of
(4.9\,$\pm$\,2.6)\,$\times$10$^{14}$~M$_{\odot}$ within 600~kpc, and
(4.6\,$\pm$\,1.9)\,$\times$10$^{14}$~M$_{\odot}$ within a radius of
$\sim$1~Mpc, where the uncertainty is computed by a jackknife sampling
following the prescription in \citet{Beers1990}. These values are consistent
with each other and with the mass estimated from the lens model to within
the stated uncertainties. If the hint of an increase in the mass with
increasing radius turns out to be real, then we will have a velocity
indicator that this system that is not virialized. We refer also to
\citet{Ferragamo2020} for a discussion on the biases inherent to estimating
masses based on velocity dispersions. We refer to \S8 for a discussion of
the cluster kinematics, and to \cite{Golovich2017} for a review of the
pitfalls entailed with the approach to detect some mergers by the
non-Gaussian shapes of their velocities alone.

\section{{Revised Lens Model}} \label{sec:LTM}

We use the well-tested light traces mass (LTM) approach to construct the 2D
projected mass map as is used in F19 \citep{Zitrin2009,Zitrin2015}. { This
semi-parametric method assumes the galaxies are tracers of the dark matter, which for the case of G165 was shown to produce results consistent with
non-parametric methods (F19)}. The inputs are the photometry and positions of
the cluster members, and the redshifts of the lens and multiply-imaged
sources. The cluster member observables inform the placement of power-law
surface density profiles which are scaled based on the galaxy
brightnesses. The superposition of these density profiles is smoothed with a
Gaussian kernel to approximate the smooth dark matter distribution of the
cluster. The cluster member density profiles are assigned a relative weight
with respect to the dark matter. In turn, the total mass distribution is approximated as the sum
of the smooth dark matter and the galaxy power law surface density profiles
scaled by the galaxy relative weights. Additional parameters include values
for external shear, shear orientation, individual galaxy weights, individual
galaxy ellipticities and position angles, and arc family redshifts.

The free parameters are fit for via a MCMC minimization using thousands of
steps to produce the final model. Errors are computed by bootstrapping the
MCMC steps, sampling 100 random realizations in the MCMC and quoting the
Gaussian errors. We stress that the MCMC errors are not robust to changes in
the lensing constraints, and are purely statistical. Strong lens modeling is
an intrinsically underconstrained problem, and the statistical errors will
ultimately underestimate the true error which is dominated by the
systematics {\citep[][]{Johnson2016,Meneghetti2017,Strait2018}}.

\subsection{{Image System Designations}}
We retain the same image system
designations as in F19, with the exception that Arcs 1a,b are renamed as Arcs 1a/b,c. The Arc 1a/b
nomenclature is preferred because it acknowledges that there are two
contiguously-positioned images of this one lensed source.  To remove
ambiguity, it follows that the arc formerly named as Arc 1b is changed to
Arc 1c. {All image systems identified and used by the model are labeled in Figure \ref{fig:lens}.}

\subsection{{The G165 Model}}
We introduce herein new photometric redshift information for Arcs
2a,c and Arcs 11a,b to refine the mass map in F19 that was anchored on a
single spectroscopic redshift of a single image of a single system, Arc 1a/b.
We also update the cluster member list as described in \S5.2. We construct
the model by fixing the Arc 1a/b redshift, and updating the
constraints on Arcs 2 and 11, leaving their redshifts free to vary only
within the $1\sigma$ uncertainty of its associated photometric redshift. The
redshifts of all other arc systems (systems 3-10) are left to be optimized
by the model without constraints. We fix the weights of all cluster members
within the LTM algorithm on the basis of their brightnesses, with the
exception of the BCG, whose weight is left free to be optimized by the
model. The new photometric and spectroscopic redshifts of the cluster
members and arcs introduce additional constraints, which result in an
improvement to the lens model.

\begin{figure*}[t!]
	\centering\includegraphics[scale =0.34]{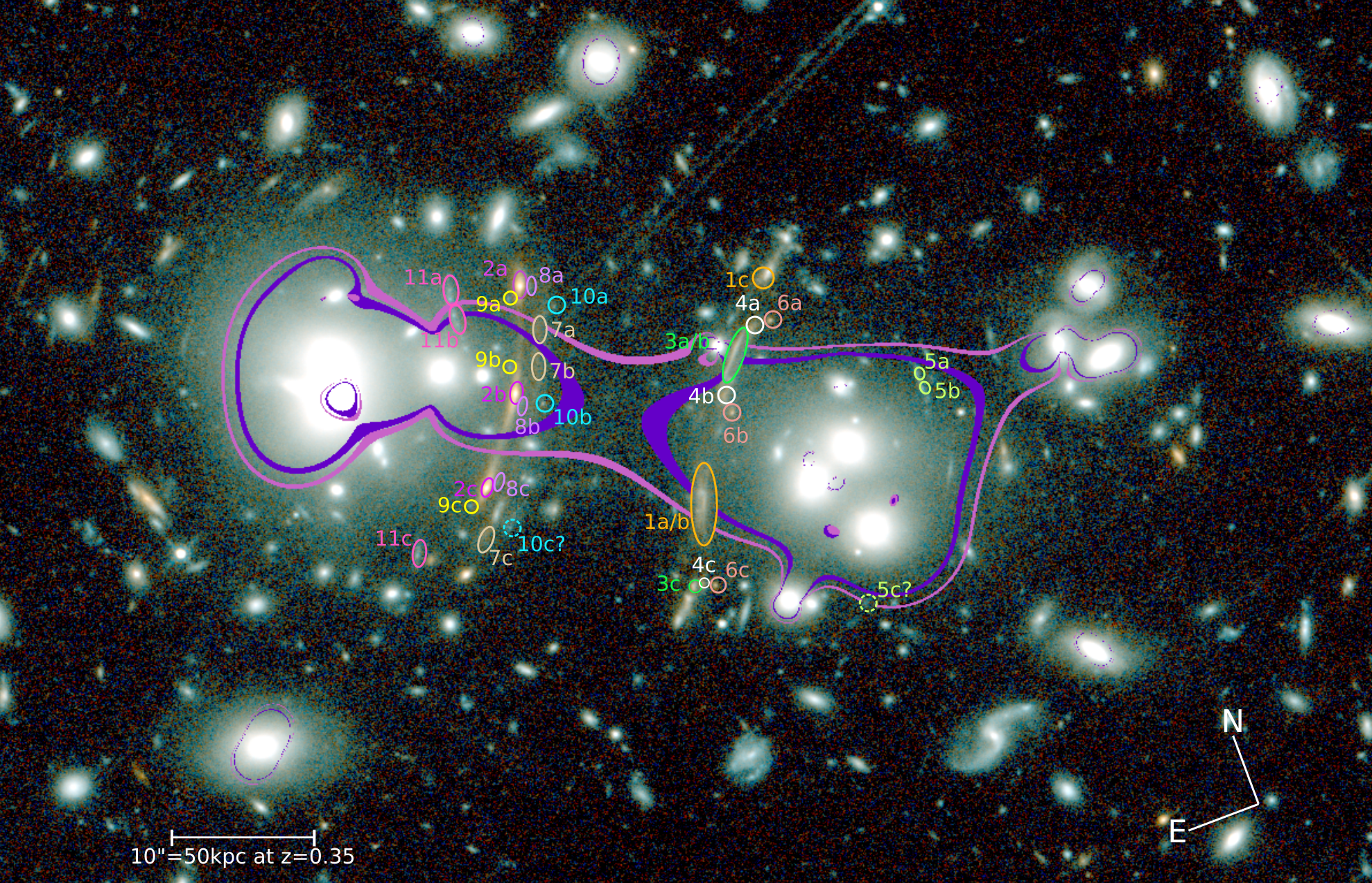}
	\caption{Composite two-color  $F110W$ and $F160W$ image showing the
$z$\,=\,2.2 (inner purple; equating to the redshift of Arc 1a/b) and the
$z$\,=\,4.8 (outer pink; equating to the redshift of Arcs 11a,b) critical
curves for the LTM lens model that incorporates the new photometric redshift
estimates obtained in this study. The critical curves are determined as an
absolute magnification cutoff in log space. All images of a single
background galaxy are coded to the same color. Several arcs fold about the
critical curve ({\it e.g.,} Arcs 1, 2, 3, 5, 7, 11), making this cluster
potentially well-suited to the detection of caustic transients.}
	\label{fig:lens}
\end{figure*}

The best model, which is the one for which $\chi^2$ is minimized, is
presented in Figure \ref{fig:lens}. Critical lines are overlaid for
$z$\,=\,2.2 (equating to the redshift of Arc 1a/b) and $z$\,=\,4.8 (equating
to the redshift of Arcs 11a,b) which delineate the elongated configuration
and bimodal mass distribution. The revised model has a smaller measured area
within the critical line when compared to the initial model in F19 by
$\sim$20\,\%. We estimate the lensing mass based on the MCMC to be
$(2.36\pm0.23)\times$\,$10^{14}M_{\odot}$ within 600 kpc. For this model we
predict input image constraints to an RMS recovery angular position of
$\sim$\,0.6$^{\prime \prime}$.  The model acquires improved functionality
also for testing the positions of the image systems in F19. The image
designations of Arc 5c and Arc 10c, which were plausible but unconfirmed in
F19, remain as such in this study. At the same time, the updated model
reduces the likelihood of the Arc 9 counterimages, 9d and 9e. We do not
uncover any new image multiplicities. We discuss the high redshift
population in \S7.3.

\begin{figure}[h]
	\centering\includegraphics[scale =0.6]{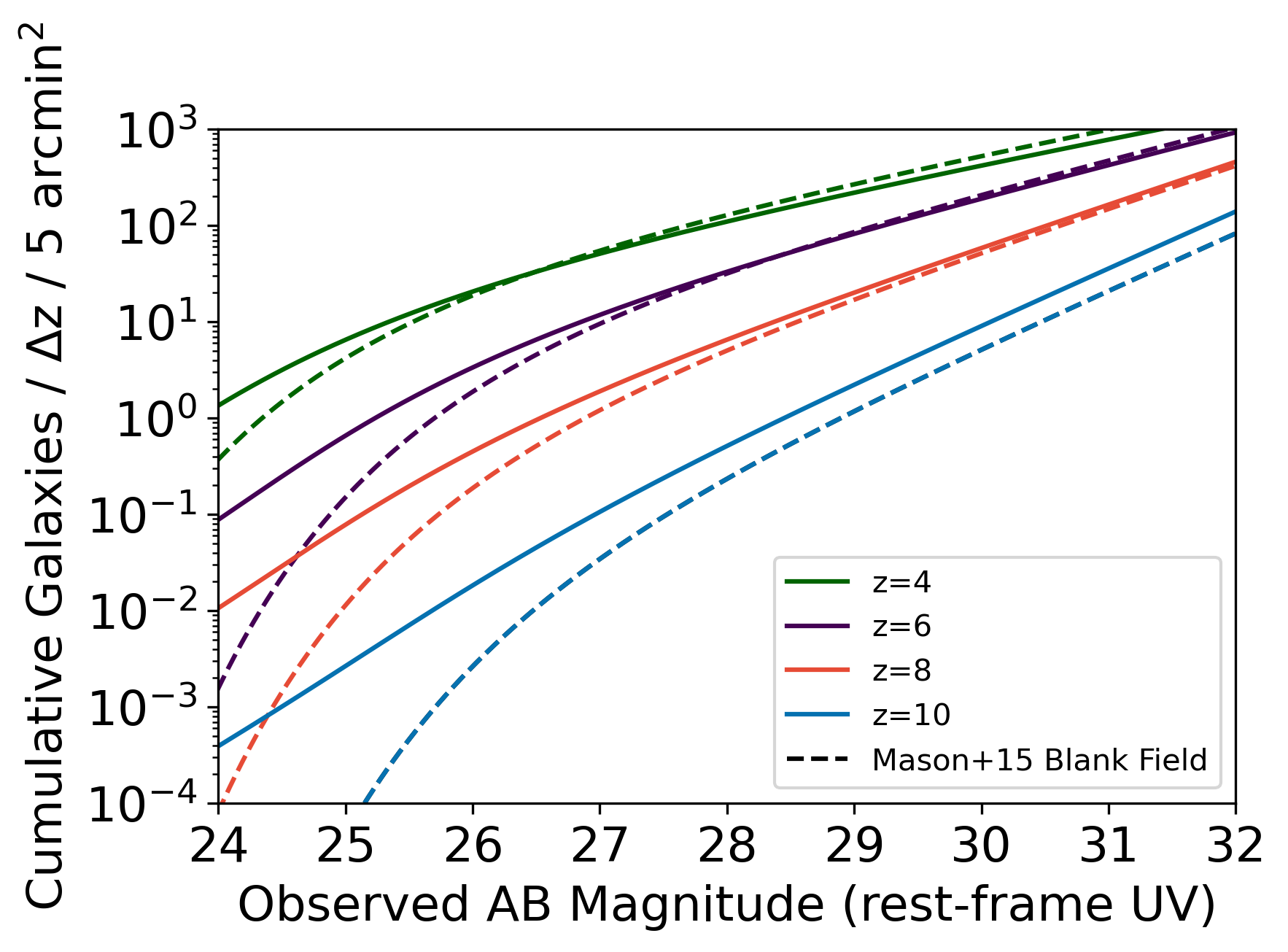}
	\caption{Cumulative number counts of lensed galaxies expected to be
detected for different redshifts given a limiting AB magnitude and assuming
a complete survey (continuous lines). We make a comparison with the blank
field luminosity functions as measured by \cite{Mason2015}, which we
extrapolate beyond the $17.5<M_{AB}<22.5$ fitting range (dashed lines). In
these models, the cumulative distribution is integrated down to the lower
limit of the fitting range of 17.5 AB mag. G165 remains a powerful lensing
cluster with a high predicted galaxy count at high redshifts in the planned
JWST observations of limiting magnitude 29 AB mag.}
	\label{fig:z9}
\end{figure}

 \begin{figure*}[t!]
	\centering\includegraphics[scale =0.88]{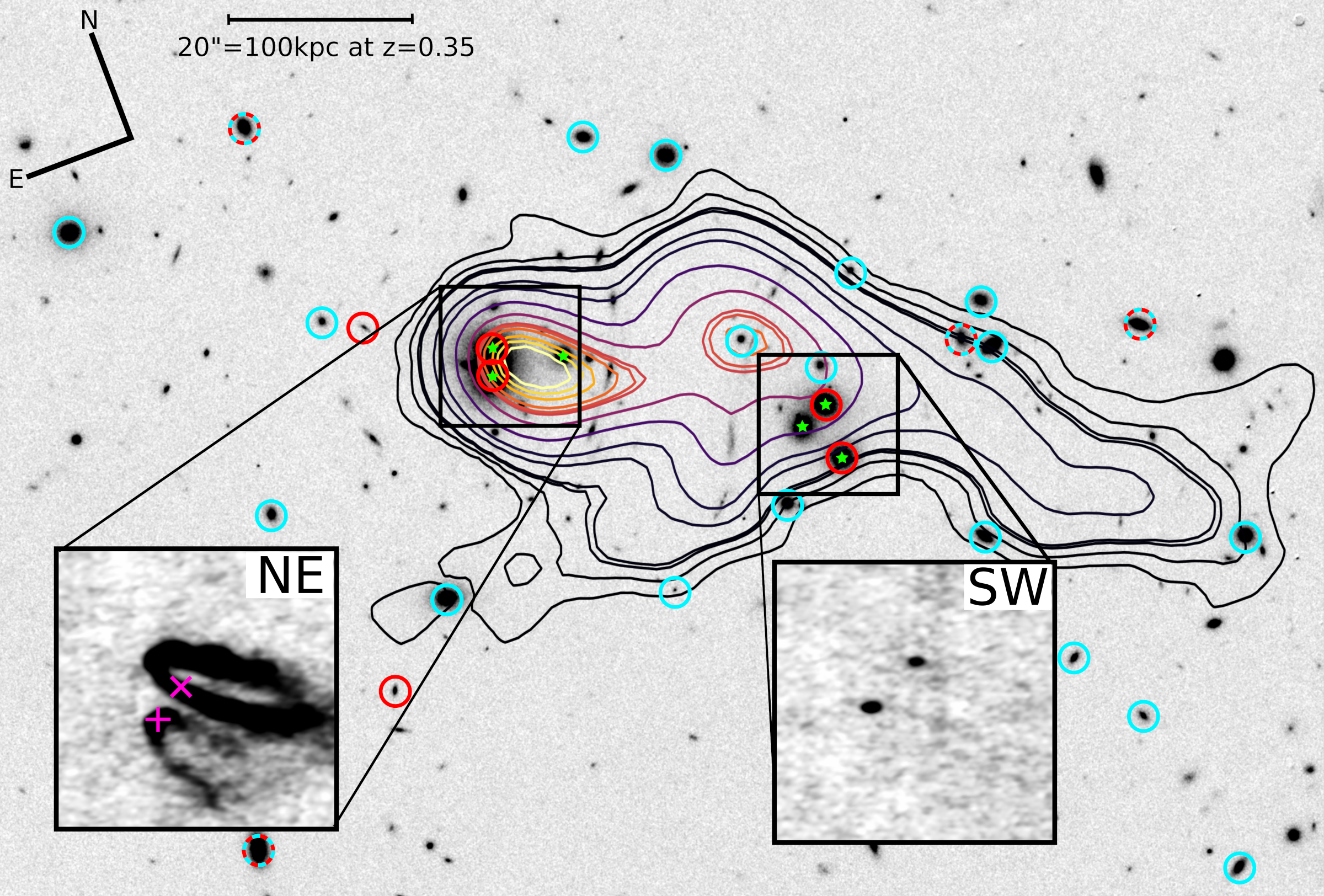}
	\caption{LUCI-Argos $K$ band image depicting spectroscopically
confirmed cluster members (0.33\,$\leq$\,$z_{spec}$\,$\leq$\,0.36, red
circles), photometrically confirmed cluster members ($0.3\leq
z_{phot}\leq0.4$, blue circles), and the set of six dominant galaxy members
(green stars).  The insets depict the VLA 6 GHz images in 15$^{\prime
\prime}$\,$\times$\,15$^{\prime \prime}$ patches centered on the SW and NE
sides, as indicated. The BCG (magenta `+'), and luminosity weighted mean of
the cluster (magenta `x'), are offset from each other.  The velocity of the
BCG is also offset from the velocity mean, all of which suggest a dynamical
disturbance. The LOFAR radio data (contours) recover the two central peaks,
and show tails extending to the southwest and to the east with a physical
extent of $\sim$\,500~kpc. Two prominent head-tail galaxies are detected in
the NE side of the VLA radio map.  These galaxies are identified with the
BCG and an LRG, {whose jet orientations suggest a post-core passage}.  Based on the available information, this cluster
appears to be undergoing a merger in a direction roughly transverse to the
LOS.}
	\label{fig:lwmc}
\end{figure*}

\subsection{Predicted Galaxy Number Counts}

G165 remains a powerful lensing cluster, with a 2D projected mass estimate
derived from our strong lensing model of $M_{600
kpc}\,=\,(2.36\pm0.23)\times$\,$10^{14}M_{\odot}$, and a $z$\,=\,9 Einstein
radius of $\theta_{E}$\,$\approx$\,15.3$^{\prime\prime}$ that is comparable
in strength to HFF cluster Abell 2744.  We apply our revised model here in
order to make predictions for the galaxy number counts extending to higher
redshifts. {This serves to place a check on the recovered photometric redshifts, as well as characterize the lensing strength implied by the model.}  Figure \ref{fig:z9} shows the cumulative number counts of lensed
galaxies observed within a 5 $\text{arcmin}^2$ FOV centered on G165 as a
function of magnitude. We opt for the \cite{Mason2015} luminosity function
to represent the blank field because it can accommodate a wide range of
redshifts, as indicated. One result is that the lensed source counts do not
win out over the field galaxy counts at all redshifts and limiting
magnitudes.  Even so, the number of lensed galaxies gains over the number of
field galaxies with increasing mean redshift.  This is because the faint end
of the luminosity function flattens with increasing redshift, such that the
loss of sources by field dilation is compensated for by the detection of
sources which rise up above the apparent limiting magnitude.

Given the limiting AB magnitude of 27 of our {\it HST} data, the {lensed
luminosity function} predicts order unity detections of $z\approx6$ and
$z\approx8$ galaxies, and dozens of $z\approx4$ galaxies. As shown in the
results of Table \ref{tab:highz}, we detect only one $z\approx6$ galaxy and
a handful of $z\approx4-5$ galaxies. {While the predictions are only rough estimates, the lower numbers of observed detections} are in part owing to
incompleteness of the catalog, as Figure \ref{fig:z9} represents an
idealized survey. Morever, although the detections made in the mixed bag of
7 bands of imaging with differing image characteristics and qualities
presented here are the best available, they are not ideal for churning out
large numbers of quality photometric redshifts. Increasing the number of
filters to improve spectral coverage at high resolution, and increasing the
depth of the 3.6\,$\mu$m and 4.5\,$\mu$m filters beyond 23 AB mag will help
to obtain a more complete photometric redshift catalog. Looking toward the
planned observations of G165 using JWST/NIRCam, a limiting magnitude of 29
AB will be achieved in 8 bands, yielding $\sim$100 sources at
$z$\,$\approx$\,4, and order unity sources at $z$\,$\approx$\,10.

\subsection{Arcs 1a/b, 1c}

The DSFG Arc 1a/b ($z=2.2$) is relatively-faint in the near-IR
($F160W_{ab}$\,=\,22.23 mag) and red ($g$\,$-$\,$i$\,=\, 0.9 AB mag), as
expected of a dust-obscured source. The high magnification factor derived
from our revised lens model of $\mu$\,$\approx25$ stretches this galaxy out
to an angular extent of $\sim$\,{5}$^{\prime \prime}$, enabling a privileged
view at the kpc scale into the interstellar medium of a high-redshift star
forming galaxy.  Arc 1a/b is detected in the rest-frame ultraviolet,
corresponding to observed $g$-band (23.4\,$\pm$\,0.1 AB mag). From this we
infer that there is at least some leakage of ultraviolet light from massive
stars.  DSFGs are highly dust-extincted, yet lensing may offer the advantage
of stretching out the light into more pixels to achieve better sampling,
which may allow the transmission of ultraviolet light. Although unresolved,
the sources of this UV starlight appear to be the star forming knots (Figure
\ref{fig:lens}).  One counter image is detected, Arc~1c, but it is faint and
drops out of detection in the three bluest filters, disallowing the
estimation of a photometric redshift. Preliminary images acquired using the
VLA detect Arc~1c, and separate out what appear to be the star forming
regions of Arc~1a/b into four knots, two each on opposite sides of the
critical curve. These results will appear in an upcoming paper (Kamieneski
et al.~2022, {\it in preparation}).

\section{{Cluster Evolutionary State}} \label{sec:cluster_ev}
\subsection{The O/IR Picture}
Our data gives clues as to the state of virialization of this cluster. We
are given: (1) the photometric catalog of 22 member galaxies which cover the
range 0.3\,$<$\,$z_{\text{phot}}$\,$<$\,0.4 out to $\sim$~600 kpc, (2) the
spectroscopic catalog of 21 cluster galaxies which covers the range
0.330\,$<$\,$z$\,$<$\,0.366 within 1 Mpc, (3) the combined catalog of the
four galaxies with both spectroscopic and photometric redshifts, and (4) the
set of six central dominant galaxies.  These six galaxies consist of three
cluster members in the northeast (NE) component, and three cluster members
in the southwest (SW) component (green stars in Figure
\ref{fig:lwmc}). Although it is useful to further subdivide the cluster
membership by galaxy activity levels and galaxy colors, such designations
are limited given the small numbers of objects.  Below we measure the
quantities available to us which were demonstrated by \citet{Rumbaugh2018}
to be correlated with cluster virialization: the offset of the luminosity
mean center from the BCG and from the velocity center.

We identify the BCG as the cluster member with the highest luminosity as
measured in $i$-band (``+" symbol in Figure \ref{fig:lwmc}. The $i$-band is
selected because it is redder than the 4000~\AA\ and Balmer breaks at the
cluster redshift. The mean projected centroid of the cluster member
luminosities is computed, from which we obtain the luminosity weighted mean
center (LWMC; ``x" symbol in Figure \ref{fig:lwmc}). The two centers are
offset by 3.3 arcseconds, equating to 16.5~kpc.  We note that although the
$i$-band alone does not fairly represent the stellar mass, the FOV of the
near-IR filters is too narrow to cover the full cluster member catalog
(Figure \ref{fig:foot}). We do not undertake an analysis of the individual
galaxy masses, nor compute a cluster mass centroid, because the set of 22
galaxies with fitted SEDs does not include certain crucial members such as
the BCG and some of the LRGs.

The velocity histogram of the cluster members has a single broad peak.  This
suggests that the two sides are moving transverse to the line-of-sight.
Three of the four central dominant members with spectroscopic redshifts have
minimal pair-wise velocity offsets of $\apll$\,100 km s$^{-1}$ (red circles
in Figure~\ref{fig:lwmc}), providing further support that the major axis of
the cluster is aligned preferentially in the plane of the sky.
Intriguingly, the BCG is blueshifted by 2400 km~s$^{-1}$ relative to the
velocity center (Figure \ref{fig:vhist}). The BCG has all the spectroscopic
features expected of a passively-evolving elliptical galaxy, from the
prominent 4000~\AA\ and Balmer breaks, to the G-band and Balmer lines in
absorption. What is unexpected is the large velocity difference between the
BCG and the cluster's systemic velocity.  It is interesting that, despite
its designation as the BCG, there are no spectroscopically-confirmed
neighbors at a similar velocity (to within 1000~km~s$^{-1}$).  Additional
longslit spectroscopy centered on the BCG would be advantageous to search
for its entourage.

\subsection{The Radio Picture}
Our radio maps corroborate the general picture of a cluster disturbance, and
give clues as to the direction of motion of the NE and SW components. The
LOFAR LoTSS map depicts emission with a physical extent of $\sim$500~kpc
that is elongated along the major-axis of the cluster.  There are two
dominant radio peaks, one roughly centered on each of the NE and the SW
components. An elongated region of somewhat more diffuse radio emission
extends to the southwest of the SW peak with a physical extent 300~kpc, and
a shelf-like emission feature is detected protruding to the east. A
distinctive comet-like morphology of the NE peak trails off into the
southwesterly direction.

At the higher resolution of the VLA 6 GHz radio map, the NE peak separates
out into two head-tail galaxies: one corresponding to the BCG and the other
to an LRG (NE inset in Figure \ref{fig:lwmc}). The VLA map also detects two
LRGs in the SW side as more compact radio sources in the same pointing, and
all within the primary beam. The trails of each of the two head-tail galaxies
are associated with an active galactic nucleus in the center of each galaxy
that is ejecting a bipolar jet.  

These Narrow Angle Tails
(NATs) 

\citep[e.~g.,][]{Muller2022} are
relatively common in galaxy cluster fields \citep[{\it
e.~g.},][]{Malavasi2016, Garon2019}. Interestingly, in the G165 field the
trails of the two NATs are also more-or-less aligned with each other, a
behavior that suggests a common ensemble jet motion.  
Given that a NAT will typically occupy the wake of its host, 
{

we infer that the radio jets are being swept back from the northeast direction by ram pressure.}

Curiously, G165 is not X-ray bright despite participating in an ongoing merger that should have shock-heated the intercluster gas to
X-ray temperatures.  {In order to maintain the lower X-ray luminosity it is tempting to speculate
that the two components may not have directly impacted each other, but instead achieved a longer distance cluster-cluster interaction.}

The near-alignment of both pairs of radio trails suggests an ensemble motion that
is not along the line-of-sight, making the transverse velocity  nonzero \citep{Gendron-Marsolais2020}.  There is not yet enough
information to measure its value, but there are some clues.  In a study
incorporating the results of NATs in both poor and rich galaxy clusters,
\citet{Venkatesan1994} find the appearance of NATs in all cases, and
estimate typical NAT velocities of $\sim$\,600~km~s$^{-1}$ to be able to
form its distinctive morphology.  In the large N-body dark-matter only
numerical simulation, the JUropa Hubble volumE
\citep[Jubilee;][]{Watson2014}, they consider examples of merging halo pairs
in the $z$\,=\,0.32 redshift slice that is well-matched to that of
G165. Although the velocity scatter is large, and the halo separations are
less well sampled at $\apll$\,0.6~Mpc, an extrapolation to smaller halo
separations indicates that typical pair-wise velocities of halo-halo mergers
are $\apg$100~km\,s$^{-1}$ \citep{Watson2014}.  In another recent study of
three HFFs, space velocities at regular intervals in the range of
500\,--\,5000 km~s$^{-1}$ are inserted to the measured velocities of the
cluster members and its deviation from the line-of-sight velocity
distribution recorded.  By this kinematical analysis, an upper limit on the
transverse velocity of 1700 km~s$^{-1}$ is allowed in order to still recover
the observed line-of-sight velocity distribution \citep{Windhorst2018}.
These studies suggest that the transverse velocity of G165 may be in the
100--1700~km~s$^{-1}$ range. X-ray observations will enable a study of the
collision properties, and improved constraints regarding the viewing angle
and transverse velocity.

\section{{Conclusions}} \label{sec:end}

{G165 is a powerful lensing cluster with a bimodal mass distribution,
rich lensing evidence, and low X-ray luminosity. Upon incorporating the new LBT/LBC
and {\it Spitzer} imaging data, photometric redshifts are enabled for three image multiplicities. New MMT/Binospec spectroscopy contributed eight additional cluster members.  These lensing constraints produced a lens model that refined the placement of the critical curve relative to the caustic-crossing arcs and 
but falls short of detecting the subhalo underlying the BCG velocity outlier. 

The detection of the four radio trails in a roughly mutual alignment suggests an impact orientation in the plane of the sky and a direction of motion in the northeast-southwest direction.
 In this scenario, the NE and SW components have already traversed each other in an event that instigated the radio activity and supplied the pressure for the cluster gas to sweep past the radio jets.  Here the low X-ray luminosity is explained by a more indirect cluster-cluster encounter of the NE and SW components.

High-resolution blue imaging is needed  to constrain the age of the merger by its rest-frame ultraviolet-optical colors,  and X-ray observations enable searches for shocks to establish the collision speed. The only way to uncover all halos participating in the cluster merger is by obtaining additional lensing evidence by deep high-resolution imaging in the planned  {\it JWST}/NIRCam PEARLS-Clusters approved program.

Ultimately, such observations offer a viable route to constrain the evolutionary state of this binary cluster that in turn contributes to our understanding of mass assembly on cluster scales.}

\acknowledgments
{We thank the referee for valuable comments
that improved the manuscript.}
We appreciate helpful discussions with Neta Bahcall, Aliza Beverage, Megan
Donahue, Matt Lehnert, James Lowenthal, Genvi\`{e}ve Soucail, Daniel Wang,
David Weinberg, and Ann Zabludoff.  Support for program {\it HST} GO-14223
was provided by NASA through a grant from the Space Telescope Science
Institute, which is operated by the Association of Universities for Research
in Astronomy, Inc., under NASA Contract NAS5-26555.

MP was funded through the NSF Graduate Research Fellowship grant No. DGE
1752814, and acknowledges the support of System76 for providing computer
equipment.  BF gratefully acknowledges the hospitality of the Institute for
Advanced Study and financial support of the Ambrose Monell Foundation and
the Bershadsky Fund during much of this work.

LD acknowledges the research grant support from the Alfred P.  Sloan
Foundation (award number FG-2021-16495). R.A.W.~was funded by NASA JWST
Interdisciplinary Scientist grants NAG5-12460, NNX14AN10G, and
80GNSSC18K0200 from NASA Goddard Space Flight Center. NF and AMB were both
funded by a UA/NASA Space Grant for Undergraduate Research. AZ acknowledges support by Grant No.~2020750 from the United States-Israel Binational Science Foundation (BSF) and Grant No.~2109066 from the United States National Science Foundation (NSF), and by the Ministry of Science \& Technology, Israel.  JMD acknowledges the support of project
PGC2018-101814-B-100 (MCIU/AEI/MINECO/FEDER, UE) Ministerio de Ciencia,
Investigaci\'on y Universidades. This project was funded by the Agencia
Estatal de Investigaci\'on, Unidad de Excelencia Mar\'ia de Maeztu,
ref.~MDM-2017-0765. This work makes use of the National Radio Astronomy
Observatory, which is a facility of the National Science Foundation operated
under cooperative agreement by Associated Universities, Inc.  We also made
use of the Large Binocular Telescope, which is overseen by an international
collaboration among institutions in the United States, Italy, and Germany.
We would like to thank the staff at the MMT, for performing the observations
in service mode.  This work has made use of data from the European Space
Agency (ESA) mission GAIA (\url{https://www.cosmos.esa.int/gaia}), processed
by the {\it Gaia} Data Processing and Analysis Consortium (DPAC,
\url{https://www.cosmos.esa.int/web/gaia/dpac/consortium}). Funding for the
DPAC has been provided by national institutions, in particular the
institutions participating in the {\it Gaia} Multilateral Agreement.

\software{{SCAMP software \citep{Bertin2002,Bertin2006}, CASA \citep{McMullin2007}, dilate \citep{DeSantis2007}, BPZ software package \citep{Benitez1999,Benitez2000,Coe2006}, astropy \citep{Astropy2013,Astropy2018}, Photutils \citep{Bradley2020}, Galfit \citep[v3;][]{Peng2010}, SExtractor \citep{Bertin1996}, Theli \citep{Erben2005,Schirmer2013}, T-PHOT\citep{Merlin2015,Merlin2016a}}}
\newpage
\bibliography{sample63}{}
\bibliographystyle{aasjournal}

\end{document}